\documentclass[a4paper,11pt]{article}
\pdfoutput=1 

\usepackage{jheppub} 
\usepackage{bbm,bm,graphicx,mathtools,color,slashed,hyperref}
\usepackage{amssymb,amsmath,amsfonts}
\usepackage{braket}

\usepackage{float}

\usepackage{subcaption}

 \usepackage{ascmac}

\newcommand{\rd}{\mathrm{d}}
\newcommand{\p}{\partial}

\newcommand{\tr}{\mathrm{tr}}
\newcommand{\im}{\mathrm{i}}

\newcommand{\rme}{\mathrm{e}}

\newcommand{\SU}{\text{SU}}
\newcommand{\SO}{\text{SO}}
\newcommand{\U}{\text{U}}
\newcommand{\ee}{\,\mathrm{e}}

\newcommand{\diag}{\operatorname{diag}} 
\newcommand{\ev}[1]{\left \langle #1 \right \rangle}
\newcommand{\abs}[1]{\left | #1 \right |}

\title{
Deconfinement-Higgs continuity in $\SU(2)$ adjoint Higgs model at finite temperature
}

\author[1]{Yui Hayashi,}
\emailAdd{yui.hayashi@yukawa.kyoto-u.ac.jp}
\affiliation[1]{
Yukawa Institute for Theoretical Physics, Kyoto University,
Kitashirakawa Oiwakecho, Sakyo-ku, Kyoto 606-8502, Japan
}

\author[1,2]{Masashi Kawahira,}
\emailAdd{masashi.kawahira@yukawa.kyoto-u.ac.jp}
\affiliation[2]{Department of Physics, Graduate School of Science, Kobe University, 1-1 Rokkodai, Kobe, Hyogo 657-8501, Japan}

\author[1,3]{Hiromasa Watanabe,}
\emailAdd{hiromasa.watanabe@keio.jp}
\affiliation[3]{Department of Physics, and Research and Education Center for Natural Sciences, Keio University, 4-1-1 Hiyoshi, Yokohama, Kanagawa 223-8521, Japan}

\preprint{YITP-25-142,\ KOBE-COSMO-25-15}

\abstract{
We study the finite-temperature phase structure of the four-dimensional
$\SU(2)$ adjoint Higgs model, focusing on a possible
\textit{deconfinement-Higgs continuity}: the conjecture that the high-temperature
deconfined phase of Yang--Mills theory and the finite-temperature Higgs
phase form a single thermodynamic phase.
We first perform a global-symmetry analysis, showing that the Higgs and
deconfined regimes are expected to share the same symmetry pattern, which
is distinct from that of the confined phase. This suggests
deconfinement--Higgs continuity, but does not exclude the possibility that
the deconfined phase and the Higgs phase are separated by a phase transition
not associated with the global symmetries.
We then perform a deformation analysis, which yields an explicit continuous
path between the ``deconfined symmetric'' and ``deconfined Higgs'' regions
in a reduced three-dimensional lattice model. 
These results
indicate that the Higgs and deconfined regimes can be continuously connected,
while the confined phase remains distinct.
}

\begin{document}

\maketitle

\section{Introduction}

Gauge theories exhibit rich phase structures, including the Coulomb phase, the Higgs phase, and the confining phase.
Understanding such a phase structure is one of the fundamental questions in quantum field theory.
The modern understanding of phases in gauge theories has significantly evolved beyond the conventional Landau paradigm, which relies on local, symmetry-breaking order parameters.
This shift is necessitated by foundational principles such as Elitzur's theorem \cite{Elitzur:1975im}, which states that local gauge symmetries cannot be spontaneously broken, thereby invalidating gauge-variant local fields as true order parameters.\footnote{
There are several works on the Higgs mechanism in a gauge-invariant manner.
A classic work is that of Fr\"ohlich-Morchio-Strocchi \cite{Frohlich:1980gj, Frohlich:1981yi}, which was revisited in \cite{Kondo:2016ywd, Kondo:2018qus}.
A recent review can be found in \cite{Maas:2017wzi}.
} 
Consequently, the classification of phases now relies on the behavior of gauge-invariant, non-local observables and the realization of global symmetries.
The recent advent of higher-form symmetries \cite{Gaiotto:2014kfa,Kapustin:2014gua} has provided a powerful and precise language for this task.

Gauge theories with scalar fields are interesting playgrounds to understand phase structures of four-dimensional gauge theories, as they possess an interesting relationship between the confining and Higgs regimes. 
The nature of this relationship depends critically on the representation of the scalar matter field under the gauge group. 
For models with scalar fields in the fundamental representation, it is well-established that the confining and Higgs regimes are analytically connected, a phenomenon known as Fradkin-Shenker continuity \cite{Fradkin:1978dv, Osterwalder:1977pc, Banks:1979fi}.

The situation is qualitatively different and far more subtle for models with scalar fields in the adjoint representation.
Since adjoint fields are neutral under the center of the gauge group, the 1-form center symmetry remains an exact symmetry of the action. This preservation allows for the possibility of a genuine thermodynamic phase transition separating the confining phase from the Higgs phase.
This makes adjoint Higgs models a richer and more complex theoretical laboratory.
Thus, we study the simplest $\SU(2)$ adjoint Higgs model, at zero and finite temperature, in this paper.

Adjoint Higgs models are not merely a theoretical curiosity; they are often building blocks of several grand unified theories (GUTs) \cite{Georgi:1974sy,Fritzsch:1974nn,Pati:1974yy}.
Furthermore, the adjoint Higgs model has a deep connection with the magnetic monopole condensation picture for the confinement mechanism.
For example, the supersymmetric version \cite{Seiberg:1994rs} provides a concrete realization of the dual superconductor picture.
In addition, phase diagrams with adjoint matter at finite temperature are of interest in the large-$N$ context \cite{Aharony:2003sx,Aharony:2005bq}, offering alternative paths to understanding the phase structure of gauge theories.
Furthermore, the adjoint Higgs model also attracts interest in the context of condensed matter physics \cite{Catumba:2024jau}.

In this paper, we focus on the simplest nontrivial example: the $\SU(2)$ adjoint Higgs model, mainly at finite temperature. 
There have been many studies on the phase structures of the adjoint Higgs models at zero and finite temperature \cite{Lang:1981qg, Brower:1982yn, Karsch:1983ps, Baier:1986ni, Baier:1988sc, Nishimura:2011md, Shibata:2023hfy, Ikeda:2023lwk}.\footnote{One should be careful to say ``finite-temperature'' in the lattice model.
Here, we only mean the phase diagram on a lattice with finite temporal sites, and we do not precisely determine the temperature in the continuum limit.}
In particular, Ref.~\cite{Karsch:1983ps}  conducted a lattice Monte Carlo simulation on a $6^3\times3$ lattice.
While they found a clear signal for the deconfining phase transition, they only suggested --- but did not unambiguously observe---  a phase transition between the high-temperature deconfined regime (``deconfined symmetric'') and the Higgs regime (``deconfined Higgs'').
More recently, in \cite{Nishimura:2011md} they studied the $\SU(2)$ adjoint Higgs model on 
$\mathbb{R}^3 \times S^1$ using semiclassical methods enabled by a ``center-stabilizing'' deformation of the action.
Their analysis claims a rich structure with four distinct phases -- confined, deconfined, Higgs, and a ``mixed confined'' phase -- sharply distinguished by their patterns of global symmetry breaking.

In this work, we revisit these studies from a new perspective.
The main proposal of this paper is the possibility of a \textit{deconfinement-Higgs continuity}: the notion that the high-temperature deconfined phase of Yang-Mills theory and the finite-temperature Higgs phase may not be distinct thermodynamic phases, but rather form a single, continuously connected phase.

The rest of this paper is organized as follows.
In Section~\ref{sec:our_motivation_and_proposal}, 
we introduce the $\SU(2)$ adjoint Higgs model and propose a natural scenario for the phase diagram.
In Section~\ref{sec:global_symmetry_analysis}, we examine the behavior of the global symmetries in our model.
In Section~\ref{sec:center_destabilization_analysis}, we explicitly construct the path connecting the deconfined phase and the Higgs phase.
Section~\ref{sec:summary_and_discussion} is devoted to a summary and future directions.

\section{Our motivation and proposal}
\label{sec:our_motivation_and_proposal}

\subsection{Setup}

In this paper, we consider the coupled system of an $\SU(2)$ gauge field\footnote{
Some references assert that, when coupled to an adjoint Higgs field, the relevant gauge group is $\SO(3)$. However, as noted in \cite{Aharony:2013hda}, it can in fact be coupled to an $\SU(2)$ gauge field. This subtle difference between $\SO(3)$ and $\SU(2)$ affects the normalization of topological operators for emergent symmetries in Section \ref{subsec:Analysis_at_zero_temperature_(1)}, and thus requires care.
} 
$a=a_\mu^i \sigma^i\rd x^\mu$
and an adjoint Higgs field $\phi=\phi^i \sigma^i$---namely, the $\SU(2)$ adjoint Higgs model,
where $\sigma^i\ (i=1,2,3\ {\rm or}\ i=x,y,z)$ are the Pauli matrices.

The action is given by
\begin{align}
&S
=
S_{\rm gauge}
+
S_{\rm Higgs}
,\label{eq:action}\\
&
S_{\rm  gauge}
=
\int_{\mathbb{R}^3\times S^1_{\beta}} 
\frac{1}{4g^2}
\tr
\left[
f\wedge \star f
\right]
,\label{eq:action_gauge}\\
&
S_{\rm Higgs}
=
\int_{\mathbb{R}^3\times S^1_{\beta}}
\rd^4 x
\left(
\frac{1}{4}
\tr
\left[
(D_\mu \phi)^2
\right]
+V(\phi)
\right),
\label{eq:action_Higgs}
\end{align}
where $f:=\rd a+\im a\wedge a$, and the covariant derivative is defined as $D_\mu \phi:=\p_\mu \phi-\im[a_\mu,\phi]$.
Furthermore, to take finite-temperature effects into account, we compactify the time direction on a circle $S^1$ with period $\beta$.
The potential\footnote{
Note that, in the case of $\SU(N)$ gauge theory, the realization of the Higgs mechanism depends on its specific form. As discussed in Appendix \ref{sec:Higgs_phase_in_SU(N)}, this form of potential is employed in grand unified theories (GUTs) as well.
} is taken to be
\begin{align}
    V(\phi)=\tr[m^2\phi^2+\lambda\phi^4].
    \label{eq:potential}
\end{align}
We assume that $m^2$ can take values ranging from $-\infty$ to $+\infty$.

\subsection{Naive phase diagram}
\label{subsec:Naive_phase_diagram}

\begin{figure}[t]
  \centering
  \includegraphics[width=0.7\textwidth]{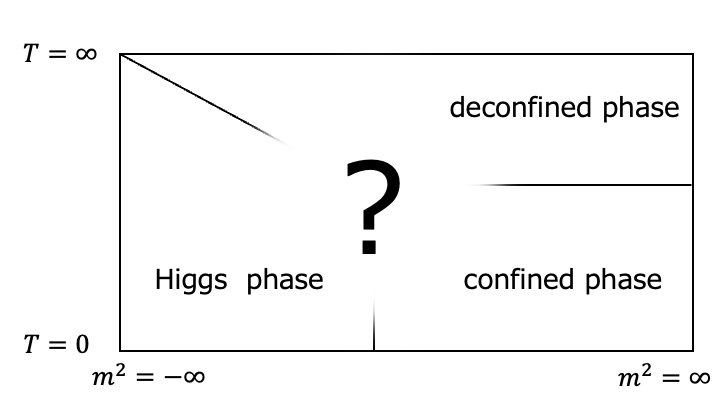}
  \caption{Naive phase diagram in the $(m^2, T)$ plane.}
  \label{fig:naive_phase_diagram}
\end{figure}

Under the above setup, we draw the phase diagram with the vertical axis $T:=1/\beta\in[0,\infty)$ and the horizontal axis $m^2\in(-\infty,\infty)$.
Let us begin with three naive observations:
\begin{itemize}
\item 
For $m^2\gg \Lambda_0^2$, the system is in the symmetric phase.\footnote{
Here $\Lambda_0$ denotes the dynamical scale of the $\SU(2)$ gauge theory.
} In this regime, it can be regarded as a finite-temperature $\SU(2)$ gauge theory, which exhibits both deconfined and confined phases
\item 
For $m^2\ll -\Lambda_0^2$, the Higgs mechanism occurs, and the theory becomes effectively a $\U(1)$ gauge theory. We refer to this phase as the Higgs phase.
\item 
According to finite-temperature perturbative calculations, it is expected that gauge symmetry is restored in the high-temperature region.
\end{itemize}
From these observations, the phase diagram in Figure \ref{fig:naive_phase_diagram} is obtained.
It should be emphasized that the Higgs mechanism is, in the first place, a perturbative concept. The Higgs mechanism can be trusted only in the regime $m^2\ll -\Lambda_0^2$.
Moreover, finite-temperature perturbative calculations are reliable only in the limit $T\gg \Lambda_0$.
Therefore, in the bulk region where both $m^2$ and $T$ take finite values, the situation is not well understood.

\subsection{Our proposal}

The purpose of this study is to uncover the bulk region. Throughout this paper, we propose that \textit{deconfinement–Higgs continuity} is a possible scenario in the bulk region.
\begin{figure}[t]
  \centering
  \includegraphics[width=0.7\textwidth]{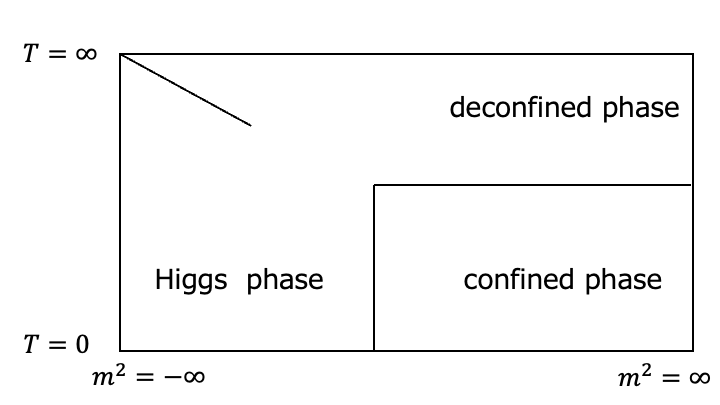}
  \caption{Our proposed phase diagram in the $(m^2, T)$ plane.}
  \label{fig:our_proposed_phase_diagram}
\end{figure}
As shown in Figure \ref{fig:our_proposed_phase_diagram}, the deconfinement–Higgs continuity is the claim that the Higgs and deconfined phases are continuously connected.

In this paper, we perform a global symmetry analysis (Section \ref{sec:global_symmetry_analysis}),
 and a center-destabilizing deformation analysis (Section \ref{sec:center_destabilization_analysis}).
These analyses are consistent with deconfinement–Higgs continuity scenario.

It should be noted, however, that the center-destabilized analysis is not performed in the continuum theory given in \eqref{eq:action}, \eqref{eq:action_gauge}, and \eqref{eq:action_Higgs}, but rather in its lattice model as introduced in Section \ref{sec:center_destabilization_analysis},  in a similar spirit to the work of Fradkin and Shenker \cite{Fradkin:1978dv}.

In addition, we present preliminary Monte Carlo results in Appendix~\ref{sec:Monte_Carlo_analysis}
as a qualitative check. These results are consistent with deconfinement--Higgs continuity.

\section{Global symmetry analysis}
\label{sec:global_symmetry_analysis}

\subsection{Global symmetries at zero temperature and finite temperature}

At zero temperature,  the model defined by \eqref{eq:action}, \eqref{eq:action_gauge}, and \eqref{eq:action_Higgs} has the global symmetries as follows.\footnote{
At first glance, the Higgs reflection might seem to be a gauge transformation. However, since it acts only on the Higgs field without transforming the gauge field, it is in fact a global transformation. 
This can be seen explicitly from the fact that, although ${\rm tr}(\phi f)$ is gauge-invariant, it carries the $(\mathbb{Z}_2^{[0]})^{4\rd}_{\rm Higgs}$ charge.
}
\begin{align}
\begin{tabular}{|c|c|}
\hline
global symmetry & transformation \\
\hline
Higgs reflection symmetry 
$(\mathbb{Z}_2^{[0]})^{4\rd}_{\rm Higgs}$  
& 
$\phi\mapsto -\phi$  
\\
center symmetry
$(\mathbb{Z}_2^{[1]})^{4\rd}_{\rm center}$  
& 
$\tr (W)\mapsto -\tr (W)$
\\
\hline
\end{tabular}
\label{tab:symmetries_at_zero_temperature}
\end{align}
\noindent{}
Here,
$\tr (W)$
denotes the Wilson loop operator, defined as
\begin{align}
W(\gamma)
:=
\exp
\left(
\im\oint_{\gamma}
a_\mu^i\sigma^i\rd x^\mu
\right).
\end{align}
It carries the charge of $(\mathbb{Z}_2^{[1]})^{\rm 4d}_{\rm center}$. 
Consequently,  if $(\mathbb{Z}_2^{[1]})^{\rm 4d}_{\rm center}$ is unbroken, 
$\tr(W)$ obeys
\begin{align}
&\lim_{|\gamma|\to\infty}
\langle 
\tr(W(\gamma))
\rangle
=
0
\end{align}
corresponding to the area law. 
Conversely, if it is spontaneously broken, then
\begin{align}
&\lim_{|\gamma|\to\infty}
\langle 
\tr(W(\gamma))
\rangle
\neq
0
\end{align}
which corresponds to the perimeter law.

At finite temperature, the temporal direction is compactified with period $\beta$.
Since the phase structure is defined in the infrared limit, it is governed by a three-dimensional effective theory at scales much larger than $\beta$.
\begin{align}
\begin{tabular}{|c|c|}
\hline
global symmetry & transformation \\
\hline
Higgs reflection symmetry 
$(\mathbb{Z}_2^{[0]})^{3\rd}_{\rm Higgs}$  
& 
$\phi\mapsto -\phi$  
\\
spatial center symmetry
$(\mathbb{Z}_2^{[1]})^{3\rd}_{\rm center}$  
& 
$\tr (W_{\rm spatial})\mapsto -\tr (W_{\rm spatial})$
\\
temporal center symmetry
$(\mathbb{Z}_2^{[0]})^{3\rd}_{\rm center}$
&
$\tr (P)\mapsto -\tr (P)$     \\
\hline
\end{tabular}
\end{align}
\label{tab:symmetries_at_finite_temperature}
\noindent{}
Here,
$\tr (W_{\rm spatial})$
denotes the spatial Wilson loop, defined as
\begin{align}
W_{\rm spatial}(\gamma)
:=
\exp
\left(
\im\oint_{{\rm spatial\ loop}\ \gamma}
a_\mu^i\sigma^i\rd x^\mu
\right).
\end{align}
In addition, $\tr (P)$ denotes the Polyakov loop, defined as
\begin{align}
P
:=
\exp
\left(
\im\oint_{{\rm temporal\ loop}}
a_4^i\sigma^i\rd x^4
\right).
\end{align}

\subsection{The behavior of global symmetries}
\label{subsec:The_behavior_of_global_symmetries}

\paragraph{Symmetric region:}  
When $m^2\gg \Lambda_0^2$, the pattern of global symmetries is straightforward.
In this regime, the system reduces to an $\SU(2)$ gauge field coupled to an ordinary scalar field, and thus both $(\mathbb{Z}_2^{[0]})^{3\rd}_{\rm Higgs}$ and $(\mathbb{Z}_2^{[1]})^{3\rd}_{\rm center}$ remain unbroken. 
The temporal center symmetry $(\mathbb{Z}_2^{[0]})^{3\rd}_{\rm center}$, on the other hand, is preserved in the confined phase but spontaneously broken in the deconfined phase.
\begin{figure}[t]
  \centering
  \includegraphics[width=0.7\textwidth]{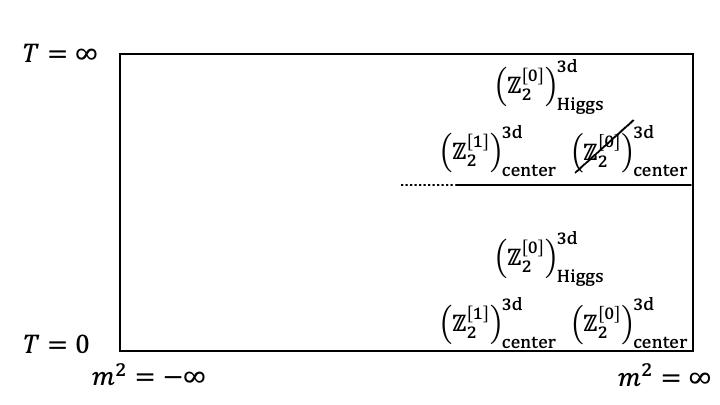}
  \caption{Global symmetry in $m^2\gg \Lambda_0^2$ region.}
  \label{fig:global_symmetry_in_m^2_gg_0}
\end{figure}

\begin{figure}[t]
  \centering
  \includegraphics[width=0.7\textwidth]{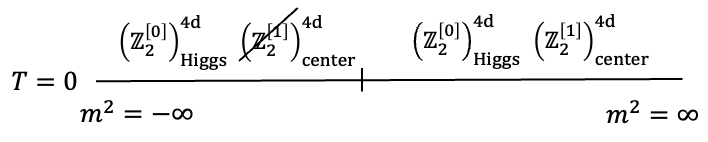}
  \caption{Global symmetry on $ T=0$ line.}
  \label{fig:global_symmetry_on_T_=_0}
\end{figure}

\paragraph{Zero-temperature line:}
The behavior of the global symmetries is as shown in Figure~\ref{fig:global_symmetry_on_T_=_0}. In the region where $(\mathbb{Z}_2^{[1]})^{4\rd}_{\rm center}$ is broken, there is expected to exist a massless photon degree of freedom, as a natural scenario. Hence, this region is identified with the familiar Higgs phase. 
These points are explained in Sections \ref{subsec:Analysis_at_zero_temperature_(1)} and \ref{subsec:Analysis_at_zero_temperature_(2)}.

\begin{figure}[t]
  \centering
  \includegraphics[width=0.7\textwidth]{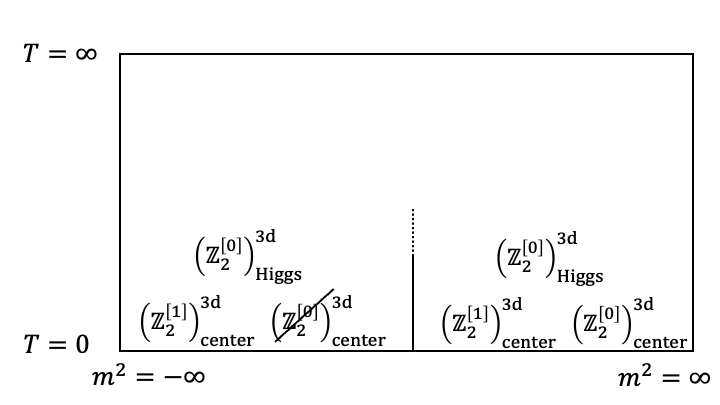}
  \caption{Global symmetry in $T\neq 0$ region.}
  \label{fig:global_symmetry_on_T_neq_0}
  \captionsetup{skip=0pt}
\end{figure}

\paragraph{Finite-temperature region:}
By incorporating finite-temperature corrections into the above situation, we obtain a minimal scenario for the phase structure, as illustrated in Figure~\ref{fig:global_symmetry_on_T_neq_0}. 
Note that while $(\mathbb{Z}_2^{[1]})^{4\rd}_{\rm center}$ is broken in Figure~\ref{fig:global_symmetry_on_T_=_0},
$(\mathbb{Z}_2^{[1]})^{3\rd}_{\rm center}$  
is expected to be 
restored in Figure~\ref{fig:global_symmetry_on_T_neq_0}. This is because the effective three-dimensional theory contains dynamical monopoles.
Further details are given in Section \ref{subsec:Analysis_at_finite_temperature}.

Note that there is a subtlety concerning Figures~\ref{fig:global_symmetry_on_T_=_0} and \ref{fig:global_symmetry_on_T_neq_0}.
In the Higgs phase, we have assumed that both $(\mathbb{Z}_2^{[0]})_{\rm Higgs}^{\rm 4d}$ and $(\mathbb{Z}_2^{[0]})_{\rm Higgs}^{\rm 3d}$ 
 remain unbroken.
As discussed in Sections~\ref{subsec:Analysis_at_zero_temperature_(1)}, \ref{subsec:Analysis_at_zero_temperature_(2)}, and \ref{subsec:Analysis_at_finite_temperature}, however, this should be regarded only as a natural scenario.
We revisit this issue in Section~\ref{subsec:remarks}.
To refine our understanding of this point, we perform further analyses in Sections~\ref{sec:center_destabilization_analysis}.

\begin{figure}[t]
  \centering
  \includegraphics[width=0.7\textwidth]{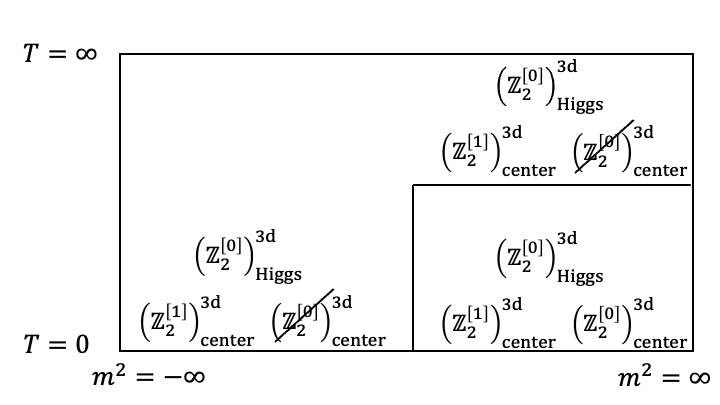}
  \caption{Global symmetry in the $(m^2, T)$ plane.}
  \label{fig:global_symmetry}
  \captionsetup{skip=0pt}
\end{figure}

By combining the above considerations, we arrive at Figure~\ref{fig:global_symmetry}. As seen in this figure, from the viewpoint of global symmetries, we expect that the Higgs and deconfined phases are indistinguishable, while the confined phase is distinct. 
The former is the broken phase of $(\mathbb{Z}_2^{[0]})^{3\rd}_{\rm center}$, whereas the latter is the symmetric phase.
Therefore, there must be a phase transition between them, and the order parameter is the Polyakov loop.

However, from this argument one cannot conclude that there is no phase transition between the Higgs and deconfined phases. An analysis based on global symmetries can only show the existence of a phase transition between certain phases, but not its absence. To establish the absence of a phase transition, additional analyses are required.

\subsection{Deep Higgs limit analysis at zero temperature}
\label{subsec:Analysis_at_zero_temperature_(1)}

Here we consider the deep Higgs limit $m^2\to-\infty$ at zero temperature. 
In this case, the Higgs mechanism takes place, and the theory is rigorously guaranteed to reduce to a $\U(1)$ gauge theory.
Consequently, the center (or electric) 1-form symmetry (or electric 1-form symmetry), $(\U(1)^{[1]})^{4\rd}_{\rm ele}$, and its dual (or magnetic) 1-form symmetry, $(\U(1)^{[1]})^{4\rd}_{\rm mag}$, emerge.

This can be also understood explicitly by focusing on the following operators:
\begin{align}
&U^{\rm ele}_{\alpha}(\Sigma):=\exp\left(\frac{2\im\alpha}{g^2} \oint_\Sigma \star f_{\U(1)}\right),
\label{eq:U_ele}\\
&U^{\rm mag}_\alpha(\Sigma):=\exp\left(\frac{\im\alpha}{2\pi} \oint_\Sigma  f_{\U(1)}\right),
\label{eq:U_mag}\\
&f_{\U(1)}:=\frac{1}{2v}\tr(\phi f),
\label{eq:f_U(1)}
\end{align}
where $v=\sqrt{-m^2/2\lambda}$, and $\Sigma$ denotes a two-dimensional closed surface.
Here, the factor of two in the expression \eqref{eq:f_U(1)} arises from the fact that we are dealing with an $\SU(2)$ gauge theory rather than an $\SO(3)$ gauge theory. Furthermore, in the deep Higgs limit $m^2\to-\infty$, the scalar field $\phi$ is trapped at the bottom of the potential and thus freezes, losing its dynamics. As a consequence, the gauge-invariant operators \eqref{eq:U_ele}, \eqref{eq:U_mag} acquire topological nature. In other words, $(\U(1)^{[1]})^{4\rd}_{\rm ele}$ and $(\U(1)^{[1]})^{4\rd}_{\rm mag}$ symmetries emerge.

Recalling that the theory possesses the center symmetry $(\mathbb{Z}_2^{[1]})^{4\rd}_{\rm center}$, one can interpret the above discussion as its enhancement to $(\U(1)^{[1]})^{4\rd}_{\rm ele}$ in the deep Higgs limit.
\begin{align}
(\mathbb{Z}_2^{[1]})^{4\rd}_{\rm center}
\xrightarrow{\rm enhancement}
(\U(1)^{[1]})^{4\rd}_{\rm ele},
\ \ \ 
{\rm as}
\ \ \ 
m^2\to-\infty.
\label{eq:Z_2_to_U(1)_enhancement}
\end{align}

This $(\U(1)^{[1]})^{4\rd}_{\rm ele}$ symmetry is expected to be spontaneously broken, because there exists a mixed 't~Hooft anomaly between $(\U(1)^{[1]})^{4\rd}_{\rm ele}$ and $(\U(1)^{[1]})^{4\rd}_{\rm mag}$.
\footnote{
This 't~Hooft anomaly is the same as that in four-dimensional Maxwell theory.
}
When $(\U(1)^{[1]})^{4\rd}_{\rm ele}$ is completely broken:
\begin{align}
(\U(1)^{[1]})^{4\rd}_{\rm ele}
\xrightarrow{\rm SSB}
\{1\}, 
\label{eq:U(1)_to_1_SSB}
\end{align}
this 't~Hooft anomaly matching condition is satisfied.
The spontaneous breaking of $(\U(1)^{[1]})^{4\rd}_{\rm ele}$ is also physically natural, since the associated Nambu–Goldstone boson is nothing but the photon. In other words, in the deep Higgs limit, the system is in the Coulomb phase.

\begin{figure}[t]
  \centering
  \includegraphics[width=0.7\textwidth]{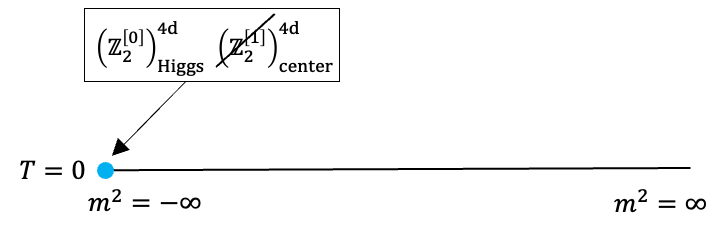}
  \caption{Global symmetry in the deep Higgs limit.}
  \label{fig:deep_Higgs_limit}
  \captionsetup{skip=0pt}
\end{figure}

To summarize the above discussion, in the deep Higgs limit,  \eqref{eq:Z_2_to_U(1)_enhancement} and \eqref{eq:U(1)_to_1_SSB} imply that
\begin{align}
(\mathbb{Z}_2^{[1]})^{4\rd}_{\rm center}
\xrightarrow{\rm SSB}
\{1\}.
\label{eq:Z_2_to_1}
\end{align}
Moreover, when this theory is viewed as a photon theory, the Higgs reflection: $\phi\mapsto -\phi$
corresponds to charge conjugation. Therefore, it is natural to expect that $(\mathbb{Z}_2^{[0]})^{4\rd}_{\rm Higgs}$ remains unbroken. 
In summary, the behavior of the symmetries at the blue point in  Figure~\ref{fig:deep_Higgs_limit} is obtained.

\subsection{Robustness  of emergent symmetries at zero temperature}
\label{subsec:Analysis_at_zero_temperature_(2)}

In the previous section, we discussed that the Coulomb-like Higgs phase emerges in the deep Higgs limit. Here, we argue that the region of this Coulomb-like Higgs phase extends to the regime of finite $m^2$.

The Coulomb phase is robust against local perturbations. This is because adding any local $\U(1)$-gauge-invariant term to the photon Lagrangian leaves the system in the Coulomb phase.
In this setup, we have non-local perturbations such as magnetic monopoles and massive gauge bosons.
However, when $-m^2$ is sufficiently large, those can be seen as local perturbations in the low-energy effective field theory.
As a result, the Coulomb phase is expected to extend outside the deep Higgs limit: $m^2\to-\infty$.

Strictly speaking, the extension of this Coulomb phase is only an expectation and has not been proven. However, in general, Nambu–Goldstone phases associated with higher-form symmetries, unlike those of 0-form symmetries, are robust against local perturbations, and in many examples it has been confirmed that the region of the Nambu–Goldstone phase extends \cite{Verresen:2022mcr}.

If we accept the above discussion, then, since \eqref{eq:Z_2_to_U(1)_enhancement} and \eqref{eq:U(1)_to_1_SSB} also hold in this Coulomb phase, the $(\mathbb{Z}_2^{[1]})_{\rm center}^{4\rd}$ symmetry
is spontaneously broken:
\begin{align}
(\mathbb{Z}_2^{[1]})^{4\rd}_{\rm center}
\xrightarrow{\rm SSB}
\{1\}.
\end{align}
Likewise, for exactly the same reason as in the previous section, 
we expect that $(\mathbb{Z}_2^{[0]})^{4\rd}_{\rm Higgs}$ remains unbroken.
Therefore, we obtain Figure~\ref{fig:global_symmetry_on_T_=_0}.

\subsection{Analysis at finite temperature}
\label{subsec:Analysis_at_finite_temperature}

From the above analysis, we have seen that the Higgs phase at zero temperature can be effectively described by a $\U(1)$ gauge theory. Let us now incorporate finite-temperature corrections.

In this case, the $(\mathbb{Z}_2^{[1]})_{\rm center}^{3\rd}$ symmetry is restored. There are two ways to understand this. 
First, since dynamical monopoles are present in this system, confinement arises according to Polyakov's argument~\cite{Polyakov:1976fu}.\footnote{
We can see it in terms of 't Hooft anomaly matching condition.
In this setup, the dual magnetic symmetry is explicitly broken, then there is not a mixed anomaly between electric and magnetic symmetries.
Furthermore, this magnetic symmetry is 0-form rather than 1-form, so the robustness argument does not apply.
} 
Second, a three-dimensional $\U(1)$ gauge theory exhibits logarithmic confinement even if the monopoles do not exist.

Moreover, when regarded as a finite-temperature $\U(1)$ gauge theory, this system is in the temporal Coulomb phase, and therefore the $(\mathbb{Z}_2^{[0]})_{\rm center}^{3\rd}$ symmetry is broken.
Furthermore, since the Higgs reflection corresponds to charge conjugation, the $(\mathbb{Z}_2^{[0]})_{\rm Higgs}^{3\rd}$ symmetry is expected to remain unbroken.
Thus we obtain Figure~\ref{fig:global_symmetry_on_T_neq_0}.

\subsection{Remarks on global symmetry analysis}
\label{subsec:remarks}

Some remarks are required regarding the content of this section.
The behavior of the three global symmetries of this system---
$(\mathbb{Z}_2^{[0]})^{\rm 3d}_{\rm Higgs}$, $(\mathbb{Z}_2^{[0]})^{\rm 3d}_{\rm center}$, and $(\mathbb{Z}_2^{[1]})^{\rm 3d}_{\rm center}$---
is shown in Figure~\ref{fig:global_symmetry}.
For the deconfined and confined phases in this figure, there is no room for doubt.

However, in the Higgs phase, the behavior of $(\mathbb{Z}_2^{[0]})^{\rm 3d}_{\rm Higgs}$ relies on several expectations and plausible reasoning.
Moreover, even if these expectations and reasoning are valid, this alone does not guarantee that the deconfined and Higgs phases are continuously connected as stated in Section~\ref{subsec:The_behavior_of_global_symmetries}.
Therefore, ultimately, unless the dynamics is fully understood, one cannot rigorously establish the continuity between the deconfined and Higgs phases.

In Section~\ref{sec:center_destabilization_analysis}, we analyze the system under a center-destabilizing deformation.
In this analysis, the temporal dynamics of the gauge field is fixed, and only the remaining dynamics is taken into account.
As a result, we find an explicit path connecting the deconfined and Higgs phases.

\section{Lattice model and center-destabilized analysis}
\label{sec:center_destabilization_analysis}

From this section, we consider a lattice analog of the finite-temperature adjoint Higgs model.
The question of the deconfinement-Higgs continuity is translated into its lattice version.
We also argue that, under the ``center-destabilizing'' deformation favoring $P= \pm I$, the deconfined phase and Higgs phase are continuously connected, in the same manner as the argument for the three-dimensional adjoint Higgs-confinement continuity.

\subsection{Map to a lattice model}

For the rest of this paper, we investigate the phase diagram of a lattice model designed to capture the essential aspects of the finite-temperature adjoint Higgs model. 
Let us consider the following lattice action (in the notation $x:= (\vec{x},t) \in \Lambda_\mathrm{lattice}$):
\begin{equation}
S[U,\phi] 
= 
\frac{\beta}{2} S_\Box[U] 
+
S_\mathrm{H}[U,\varphi]
,
\label{eq:S_latt_naive}
\end{equation}
where $S_\Box[U]$ is the Wilson plaquette action given by
\begin{align}
S_\Box[U]
:= -
\sum_{x,\mu\neq \nu}
\tr
\left[
U_\mu(x)
U_\nu(x+\hat{e}_\mu)
U^\dagger_\mu(x+\hat{e}_\nu)
U^\dagger_\nu(x)
\right],
\end{align}
and 
\begin{align}
    S_\mathrm{H}[U,\varphi]
    =
    \frac{a^4}{2}\sum_{x,\mu}
    \tr\left[
        \left(
            \frac{U_\mu(x)\varphi(x+\hat{e}_\mu)U_\mu^\dagger(x)-\varphi(x)}{a}
        \right)^2
        + m^2 \varphi(x)^2
        + \lambda \varphi(x)^4
    \right].
    \label{eq:S_latt_Higgs_naive}
\end{align}
Dynamical fields are the link variables $U_\mu(x) \in \SU(2)$ and the scalar fields $\varphi(x) = \varphi^i(x)\sigma^i$ living on the site.
The Higgs part \eqref{eq:S_latt_Higgs_naive} corresponding to the continuum one \eqref{eq:potential} can be expressed as 
\begin{align}
    &
    \frac{1}{2}\sum_{x,\mu}\tr\Bigg[
        -2a^2 \varphi(x)U_\mu(x)\varphi(x+\hat{e}_\mu)U_\mu^\dagger(x)
    \notag\\
        &\hspace{60pt} +
        \left(2+a^2m^2+2a^2\lambda v^2\right)a^2\varphi(x)^2
        +
        a^4\lambda\left(\varphi(x)^2-v^2\right)^2
    \Bigg]
    + \cdots
    \\
    =&
    \sum_{x,\mu}\tr\Bigg[
        -\frac{\beta_\mathrm{H}}{2}
        \phi(x)U_\mu(x)\phi(x+\hat{e}_\mu)U_\mu^\dagger(x)
        +
        a^2 v^2 \phi(x)^2
        +
        \frac{1}{2 }a^4 \lambda v^4 \left(\phi(x)^2-1\right)^2
    \Bigg]
    + \cdots.
\end{align}
where $v:= \sqrt{-m^2/2\lambda}$, and we introduced a lattice Higgs coupling $\beta_\mathrm{H}:= 2 a^2 v^2$ and the normalized scalar field $\phi(x) = \varphi(x)/v$, and $\cdots$ represents an irrelevant constant.
Due to our interest only in the phase structure, we focus on the simple model: we drop the radial mode by imposing a condition $\sum_{i=1}^3(\phi^i)^2 = 1$.
Intuitively, this is equivalent to considering $\lambda \to \infty$.
As a result, we derive a simplified lattice model that is of interest to us\footnote{Although this action is derived as a lattice discretization in the Higgs regime, it can also mimic a heavy adjoint scalar by choosing small $\beta_{\rm H}$. When the hopping parameter $\beta_{\rm H}$ is small, the propagation of $\phi$ is strongly suppressed, thus qualitatively realizing a heavy adjoint scalar. Therefore, although we have dropped the radial mode, the action (\ref{eq:S_latt_original}) can be considered as a good lattice analog of the adjoint Higgs model.} 
\begin{equation}
S[U,\phi] 
= 
\frac{\beta}{2} S_\Box[U] 
- 
\frac{\beta_\mathrm{H}}{2} 
\sum_{x,\mu}\phi^i(x)\phi^j(x+\hat{e}_\mu)\, \operatorname{tr} \left[U_\mu(x) \sigma^i U_\mu^\dagger(x) \sigma^j\right].
\label{eq:S_latt_original}
\end{equation}
Furthermore, we take the unitary gauge to freeze the dynamics of the scalar fields:
\begin{equation}
\sum_{x,\mu}\phi^i(x)\phi^j(x+\hat{e}_\mu)\, \operatorname{tr} \left[U_\mu(x) \sigma^i U_\mu^\dagger(x) \sigma^j\right]
\to 
\sum_{x,\mu} \operatorname{tr} \left[U_\mu(x) \sigma^y U_\mu^\dagger(x) \sigma^y\right].
\end{equation}
The convention of fixing the adjoint scalar to the $y$-direction $\sigma^y (= \sigma^2)$ in the unitary gauge follows the previous lattice study~\cite{Karsch:1983ps}.
Analogous to the continuum model, the lattice model has the following symmetry: $( \mathbb{Z}_2^{[1]} )^{\rm 3d}_{\rm center} \times ( \mathbb{Z}_{2}^{[0]})^{\rm 3d}_{\rm center} \times ( \mathbb{Z}_{2}^{[0]} )^{\rm 3d}_{\rm Higgs}$.

\subsection{Schematic phase diagram of the lattice model}
\label{sec:phase_diagram_lattice_model}

To realize ``finite temperature,'' we use a lattice $N_{\rm t} \times N_{\rm s}^3$ with finite $N_{\rm t}$.
Instead of varying the physical temperature, we consider the phase diagram in the parameter space $(\beta,\beta_\mathrm{H})$ at fixed lattice size $(N_{\rm t},N_{\rm s})$.\footnote{Strictly speaking, the phase diagram is defined in the thermodynamic limit, which is characterized by $N_{\rm s} \to \infty$ with fixed $N_{\rm t}$.}
This lattice phase diagram indeed mimics the continuum one with temperature in the following sense.

\begin{figure}[t]
    \centering
    \includegraphics[width=0.4\textwidth]{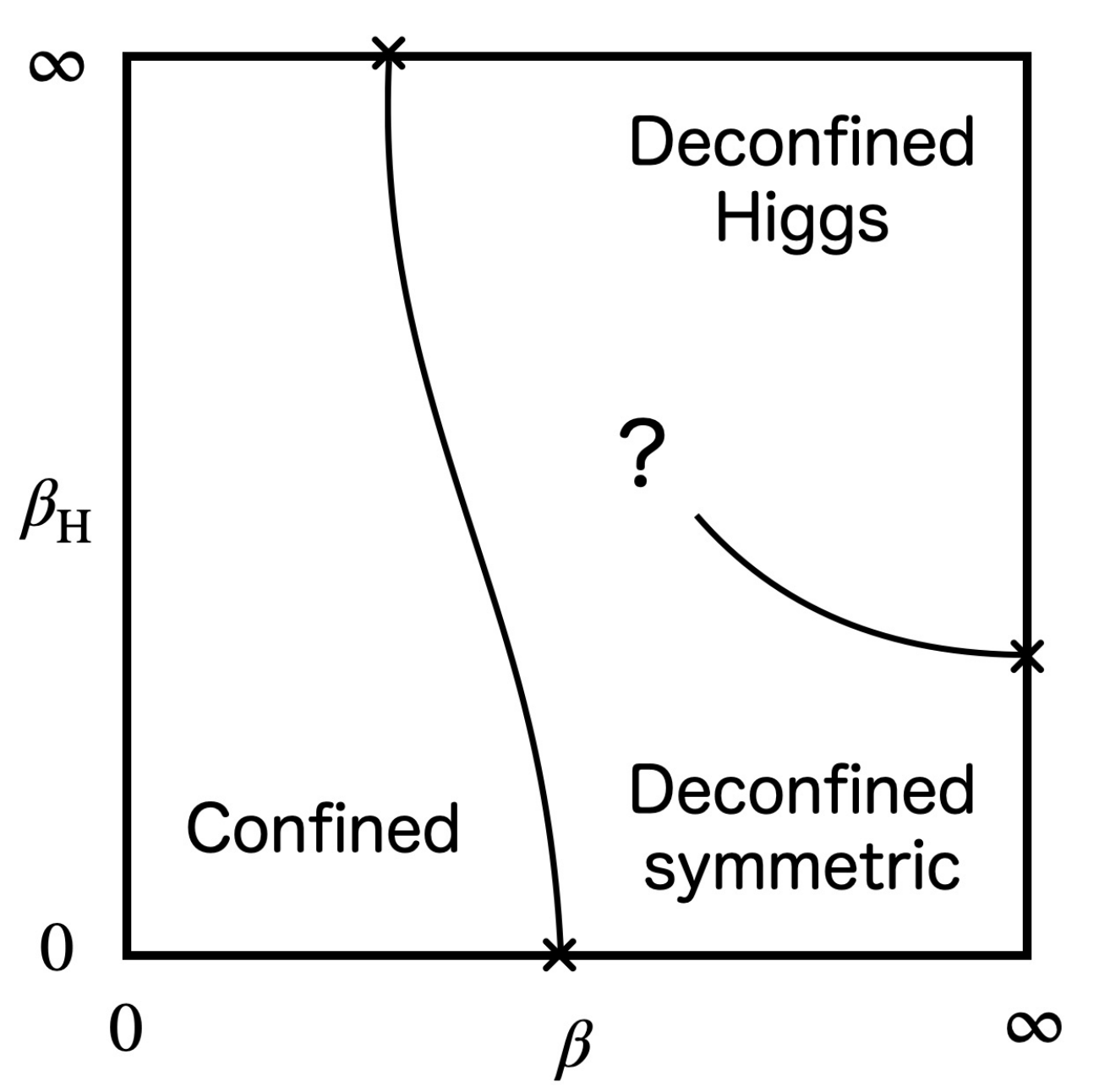}
    \caption{Schematic phase diagram of the lattice adjoint Higgs model with respect to the lattice couplings $\beta$ and $\beta_\mathrm{H}$.}
    \label{fig:PhaseDiagram_lattice}
\end{figure}

The phase structure of this lattice model was investigated nonperturbatively by lattice simulations in \cite{Brower:1982yn,Lang:1981qg,Baier:1986ni,Baier:1988sc} at zero temperature and in \cite{Karsch:1983ps} at finite temperature.
The proposed phase diagram in terms of the lattice couplings $(\beta,\beta_\mathrm{H})$ is shown in Figure~\ref{fig:PhaseDiagram_lattice}. 

Let us look at asymptotic regions corresponding to all sides of the square:
\paragraph{The top edge:}
This corresponds to the deep Higgs region, where adjoint Higgsing from $\SU(2)$ to $\U(1)$ is expected to occur.
The effective theory should be the compact $\U(1)$ gauge theory with the lattice coupling $\beta$.
The four-dimensional compact lattice $\U(1)$ gauge theory has a confinement/deconfinement phase transition at a certain value of $\beta$.
The existence of the deconfined phase has been pointed out analytically in \cite{Guth:1979gz}.
The finite-temperature compact $\U(1)$ gauge theory has been explored on the lattice in \cite{Vettorazzo:2004cr}.
Let us refer to this deconfined regime as the ``deconfined Higgs phase.''
\paragraph{The bottom edge:} This is the ordinary $\SU(2)$ pure lattice gauge theory. It exhibits the confinement/deconfinement phase transition at finite temperature, which is expected as the second-order transition from the viewpoint of universality~\cite{Svetitsky:1982gs} and by extensive lattice studies~\cite{Kuti:1980gh,Kogut:1982rt,Engels:1989fz,Engels:1990vr,Engels:1992fs,Engels:1992ke,Fingberg:1992ju,Engels:1995em,Engels:1998nv}. 
We shall refer to this deconfined regime as the ``deconfined symmetric phase.''
\paragraph{The left edge:} It corresponds to the strong-coupling limit ($\beta\to 0$) of this lattice gauge theory. This region is expected to remain confined.
\paragraph{The right edge:} In the weak-coupling limit ($\beta \to \infty$), every link variable approaches the identity. It turns out that the second term of \eqref{eq:S_latt_original} reduces to the form $-\beta_\mathrm{H} \sum_{\vec{x},m} \phi^i(\vec{x})\phi^i(\vec{x}+\hat{e}_m)$ and the ${\rm O}(3)$ Heisenberg model appears on the right edge.\footnote{
Note that, in terms of the universality class, the system is not that of the four-dimensional ${\rm O}(3)$ Heisenberg model but rather that of the three-dimensional one. (Both exhibit second-order phase transitions, but their critical exponents differ.)
} This spin model possesses a second-order phase transition at some $\beta_\mathrm{H}$. 
(See, e.g., \cite{Campostrini:2002ky,Pelissetto:2000ek}.)

\noindent{}
\noindent{}

In addition to the above discussion, a phase transition line, which we refer to as the deconfinement/Higgs transition line from now on, that extends from the right edge is conjectured by lattice simulation~\cite{Karsch:1983ps}.
The order of transition is not well-determined at present, although that work reports a negative result for a strong first-order phase transition.
As illustrated in the question-mark region of Figure~\ref{fig:PhaseDiagram_lattice}, the fate of the other endpoint also remains a mystery.

We expect that this model captures the essential features of the problem discussed in the preceding sections. In the pure Yang-Mills limit ($\beta_\mathrm{H}=0$), when the number of temporal links is fixed, a large $\beta$ corresponds roughly to the high-temperature phase, while a small $\beta$ corresponds to the low-temperature phase.
Furthermore, the hopping parameter $\beta_\mathrm{H}$ represents to the strength of Higgsing and is thus expected to correspond to the mass parameter $m^2$ in \eqref{eq:potential}.

Based on these observations, we qualitatively expect the phase diagram of this lattice model to correspond to that of the finite-temperature adjoint Higgs theory in the continuum.
Consequently, the question of whether the two deconfined phases in this lattice model are connected can be regarded as the lattice version of our original question in the continuum---namely, whether the deconfined phase and the Higgs phase are connected in Figure \ref{fig:naive_phase_diagram}.
In the remainder of this paper, we investigate the phase diagram of this lattice model.

\subsection{Center-destabilizing deformation}
\label{sec:center-destab}

The main question discussed in this paper is whether or not the deconfined Higgs phase and the deconfined symmetric phase are continuously connected.
They are separated in the weak-coupling limit, but we conjecture that these two phases are connected in the intermediate region of the phase diagram (the question-mark region in Figure~\ref{fig:PhaseDiagram_lattice}).

Here, we demonstrate that the continuity between the deconfined symmetric and deconfined Higgs phases in the large-deformation limit where the holonomy is fixed to $\pm I$\footnote{Roughly speaking, this limit corresponds to the high-temperature limit.} can be understood from the three-dimensional adjoint Higgs-confinement continuity\footnote{The phase diagram of the three-dimensional adjoint Higgs model on the lattice was investigated in Ref.~\cite{Nadkarni:1989na}.}.

In this section, we assert the existence of a path of continuous deformation connecting the deconfined symmetric and deconfined Higgs phases.
We begin with the following assumption.

\begin{screen}
\paragraph{Assumption:} The deconfined and Higgs phases do not undergo any phase transition under a deformation that destabilizes the center symmetry and fixes the holonomy to $\pm I$:
\begin{align}
    S[U,\phi] \longrightarrow S'[U,\phi] = S[U,\phi] -c \sum_{\vec{x}:~ \mathrm{spatial}} |\operatorname{tr}P(\vec{x})|^2
\end{align}    
\end{screen}
In the large-deformation limit $c \rightarrow \infty$, the Polyakov loop is fixed to $P = \pm I$, i.e., maximally center-broken vacua.\footnote{
One might suspect that the symmetry $(\mathbb{Z}_2^{[0]})_{\rm Higgs}$ is obviously unbroken
because the order parameter ${\rm tr}(P^2\phi)$ vanishes if $P=\pm I$.
However, even if the order parameter vanishes, the symmetry can generally be broken. 
(On the other hand, when the order parameter does not vanish, the symmetry must be broken.)
}

For simplicity, let us set the temporal extent be one: $N_{\rm t} = 1$.
The Polyakov loop is nothing but the single link variable $P=U_4$.
Here, we take the large-deformation limit $c \rightarrow \infty$, so we can assume that the Polyakov loop takes only $\mathbb{Z}_2$-value: $U_4 = \pm I$.

In this limit, we have the following three-dimensional lattice model: the dynamical variables are,
\begin{itemize}
\item the spatial link variable $U_m \in \SU(2)$ where $m=1,2,3$ labels spatial directions,
\item the Polyakov loop $U_4 \in \mathbb{Z}_2$, and
\item the adjoint scalar $\phi = \phi^i \sigma^i$ satisfying $\sum_{i=1}^3(\phi^i)^2 =1$.
\end{itemize}
The three-dimensional action is given by,
\begin{equation}
S_{\mathrm{3d}}[U,\phi] 
= 
\frac{\beta_{\rm s}}{2} S_\Box^{\mathrm{3d}}[U] 
- 
\frac{\beta_{\rm t}}{2} 
\sum_{\vec{x},m}\, \tr \left[U_4(\vec{x}) U_4 (\vec{x}+\hat{e}_m) \right]
- 
\frac{\beta_\mathrm{H}}{2} 
\sum_{\vec{x},m}
\, 
\operatorname{tr} \left[U_m^\dagger(\vec{x}) \phi (\vec{x}) U_m(\vec{x}) \phi(\vec{x}+\hat{e}_m) \right] ,
\end{equation}
where we treat the spatial inverse coupling $\beta_{\rm s}$ and temporal inverse coupling $\beta_{\rm t}$ as distinct parameters, and we define the three-dimensional Wilson plaquette action:
\begin{align}
     S_\Box^{\mathrm{3d}}[U] :=  -
\sum_{\vec{x}} \sum_{\substack{m,n=1,2,3 \\  m \neq n}}
\tr
\left[
U_m(\vec{x})
U_n(\vec{x}+\hat{e}_m)
U^\dagger_m(\vec{x}+\hat{e}_n)
U^\dagger_n(\vec{x})
\right].
\end{align}

\begin{figure}[t]
    \centering    \includegraphics[width=0.4\textwidth]{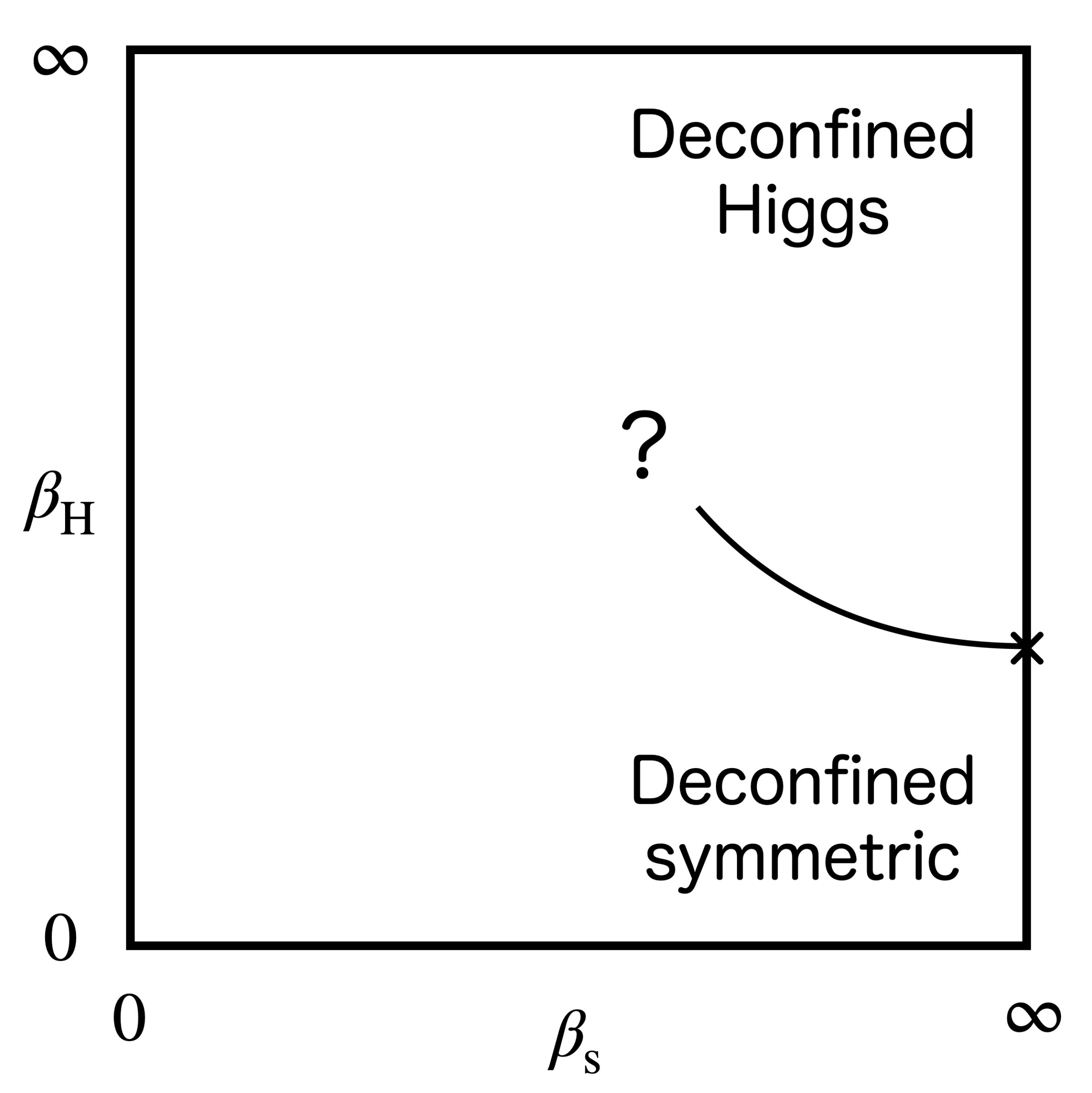}
    \caption{Schematic phase diagram of the three-dimensional lattice model with respect to the couplings $\beta_{\rm s}$ and $\beta_\mathrm{H}$.}
    \label{fig:PhaseDiagram_destabilaization_lattice}
\end{figure}

We keep $\beta_{\rm t}$ large so that the $ ( \mathbb{Z}_{2}^{[0]})^{\rm 3d}_{\mathrm{center}}$ symmetry is always broken.
Let us now consider the phase diagram on the $(\beta_{\rm s},\beta_\mathrm{H})$ plane.
By this deformation, the deconfined phase in Figure~\ref{fig:PhaseDiagram_lattice} expands, leading to a phase diagram without a confined phase.
As a result, the phase structure is expected to take the form shown in Figure~\ref{fig:PhaseDiagram_destabilaization_lattice}.

As in Figure~\ref{fig:PhaseDiagram_destabilaization_lattice}, we define the large-$\beta_{\rm s}$ and large-$\beta_\mathrm{H}$ region as the deconfined Higgs phase and the large-$\beta_{\rm s}$ and small-$\beta_\mathrm{H}$ region as the deconfined symmetric phase.
It should be noted, however, that the term ``deconfined" used here reflects the behavior of the Polyakov loop in the original four-dimensional theory.
Since the spatial Wilson loop exhibits an area law, the system is actually in a confined phase when regarded purely as a three-dimensional theory.
Hence, the continuity demonstrated below corresponds to the adjoint Higgs-confinement continuity in the three-dimensional theory as noted at the outset of Section \ref{sec:center_destabilization_analysis}.
However, to avoid confusion in terminology, we intentionally continue to use the term ``deconfined" here.

As we will see below, in this three-dimensional model one can explicitly identify a path from the deconfined Higgs phase to the deconfined symmetric phase along which no phase transition occurs.
This path is constructed by joining the following three segments:
\begin{itemize}
\item[(i)] the path from $(\beta_{\rm s},\beta_{\rm H})=(\infty,\infty)$ to $(\beta_{\rm s},\beta_{\rm H})=(0,\infty)$,
\item[(ii)] the path from $(\beta_{\rm s},\beta_{\rm H})=(0,\infty)$ to $(\beta_{\rm s},\beta_{\rm H})=(0,0)$,
\item[(iii)] the path from $(\beta_{\rm s},\beta_{\rm H})=(0,0)$ to $(\beta_{\rm s},\beta_{\rm H})=(\infty,0)$
\end{itemize}
Below, we examine these three segments one by one and show that no phase transition occurs along them.

\paragraph{(i) Deep Higgs limit $\beta_\mathrm{H} \to \infty$: }
In the deep-Higgs limit $\beta_\mathrm{H} \rightarrow  \infty$, the link variable $U_m (\vec{x}) \in \SU(2)$ is restricted to the $\U(1)$ subgroup generated by $\phi^i \sigma^i$.
Thus, the link variable $U_m (\vec{x})$ can be regarded as a $\U(1)$ variable in this limit, and we have
\begin{equation}
S_{\mathrm{3d}, \beta_\mathrm{H} \to \infty}[U,\phi] 
= 
\beta_{\rm s} S^{\mathrm{3d},\U(1)}_{\Box}[U] 
-
\frac{\beta_{\rm t}}{2} \sum_{\vec{x},m}\, 
\tr \left[U_4(\vec{x}) U_4 (\vec{x}+\hat{e}_m) \right],
\label{eq:strong_spatial_coupling_limit}
\end{equation}
where $S^{\mathrm{3d},\U(1)}_{\Box}[U] $ denotes the three-dimensional plaquette action of $\U(1)$ gauge theory.
This three-dimensional $\U(1)$ lattice gauge theory does not 
undergo a phase transition
when the coupling constant $\beta_{\rm s}$ is varied.\footnote{
This is in contrast to the phase transition that occurs along the top edge of Figure~\ref{fig:PhaseDiagram_lattice}.
The difference arises from the fact that three-dimensional Wilson-type $\U(1)$ gauge theory exhibits no phase transition as the coupling constant is varied, whereas four-dimensional theory has a phase transition.
}\footnote{
Note that the second term in \eqref{eq:strong_spatial_coupling_limit} is entirely closed within the interaction of $U_4 (x) \in \mathbb{Z}_2$ and can be seen to decouple from the rest of the dynamics.
}

\paragraph{(ii) Strong spatial coupling limit $\beta_{\rm s} \rightarrow 0$: }
The action is given by
\begin{equation}
S_{\mathrm{3d}, \beta_{\rm s} \to 0}[U,\phi] 
= 
- 
\frac{\beta_\mathrm{H}}{2} 
\sum_{\vec{x},m}\, \tr \left[U_m^\dagger(\vec{x}) \phi (\vec{x}) U_m(x) \phi(\vec{x}+\hat{e}_m) \right] - \frac{\beta_{\rm t}}{2} \sum_{\vec{x},m}\, \tr\left[U_4(\vec{x}) U_4 (\vec{x}+\hat{e}_m) \right].
\end{equation}
By reparametrizing $U_m (x) \in \SU(2)$, one can reduce it to the form
\begin{equation}
- 
\frac{\beta_\mathrm{H}}{2} 
\sum_{\vec{x},m}\, \tr \left[U_m^\dagger(\vec{x}) \phi (\vec{x}) U_m(\vec{x}) \phi(\vec{x}+\hat{e}_m) \right]  \longrightarrow  
- 
\frac{\beta_\mathrm{H}}{2} 
\sum_{\vec{x},m}\, \tr \left[U_m^\dagger(\vec{x}) \sigma_y U_m(\vec{x}) \sigma_y \right]  .
\end{equation}
This can be regarded as taking the unitary gauge.
This term factorizes completely on each link.
Hence, this theory undergoes no phase transition at all when $\beta_{\rm H}$ is varied.

\paragraph{(iii) Higgs decoupling limit $\beta_{\rm H} \rightarrow 0$: }
The action reduces to
\begin{equation}
S_{\mathrm{3d},\beta_{\rm H} \to 0}[U,\phi] 
= 
\frac{\beta_{\rm s}}{2} S_\Box^{\mathrm{3d}}[U] 
- 
\frac{\beta_{\rm t}}{2} 
\sum_{\vec{x},m}\, \tr \left[U_4(\vec{x}) U_4 (\vec{x}+\hat{e}_m) \right],
\end{equation}
This three-dimensional $\SU(2)$ lattice gauge theory does not undergo a phase transition
when the coupling constant $\beta_{\rm s}$ is varied.

The observations (i), (ii), and (iii) suggest that the deconfined symmetric and deconfined Higgs phases are continuously connected in the three-dimensional reduced model.
Hence, under the assumption presented at the beginning of this subsection, we can explicitly construct a deformation path connecting these two phases in the original four-dimensional model.

\subsection{Relation to the phase diagram in the continuum theory}

Here we discuss the phase transition line extending from the upper-left region in Figure \ref{fig:naive_phase_diagram} and Figure \ref{fig:our_proposed_phase_diagram}.
In principle, the phase transition line 
could be either a kinematical one associated with the spontaneous breaking of a global symmetry\footnote{
Especially, we care about the possibility of spontaneous breaking of $(\mathbb{Z}_2^{[0]})_{\rm Higgs}$ because the center symmetry $(\mathbb{Z}_2^{[0]})_{\rm center}$ is broken in both phases considered here.
} or a dynamical one unrelated to any global symmetry.
In this subsection, we argue that the latter interpretation, namely that it is a dynamical phase transition, is the more reasonable one.

In the high-temperature region, the Polyakov loop operator behaves as
\begin{align}
    P\sim \pm I.
\end{align}
It is close to the situation in the center destabilized lattice model.
Reminding the lattice model analysis, $(\mathbb{Z}_2^{[0]})_{\rm Higgs}$ is expected to not be broken in the deconfined symmetric phase and Higgs phase.\footnote{
The Monte Carlo calculation in Appendix \ref{sec:Monte_Carlo_analysis} supports this viewpoint.
There is also a previous study \cite{Nishimura:2011md} concerning the possible spontaneous breaking of $(\mathbb{Z}_2^{[0]})_{\rm Higgs}$. We comment on this point in Appendix \ref{sec:Nishimura-Ogilvie}.
}

Indeed, the nontrivial point of the destabilization analysis is that it supports the existence of a phase transition that is not kinematical in origin.
Thus, even in the presence of such a non-kinematical phase transition, our claim is that the Higgs phase and the symmetric deconfined phase are still not completely separated.
Therefore, the reasonable interpretation of the upper-left phase transition line in Figure \ref{fig:naive_phase_diagram} and Figure \ref{fig:our_proposed_phase_diagram}
is that it is a dynamical phase transition, and it can have an endpoint.\footnote{
Of course, this is just a minimal scenario.
It is logically possible that, once temporal dynamics is included, spontaneous breaking of $(\mathbb{Z}_2^{[0]})_{\rm Higgs}$ may occur.
If it occurs, there might exist some nontrivial intermediate phase.
}

\section{Summary and future directions}
\label{sec:summary_and_discussion}

\subsection{Summary}

In this work, we investigated the phase structure of the four-dimensional SU(2) adjoint Higgs model at finite temperature, with special emphasis on the possible deconfinement–Higgs continuity.
Our study combined 
the following complementary approaches:
\begin{enumerate}
\item 
\textbf{Global symmetry analysis}
(Section~\ref{sec:global_symmetry_analysis})
---
We give a natural scenario of the behavior of the 0-form and 1-form global symmetries at zero and finite temperature and trace their patterns of realization.
From this viewpoint the confined phase is distinguished by an unbroken temporal-center symmetry, whereas the Higgs and deconfined phases share the same pattern of broken and unbroken symmetries---suggesting, but not proving, that they can be continuously connected.
\item 
\textbf{Center-destabilized analysis}
(Section~\ref{sec:center_destabilization_analysis})
---
By introducing a deformation that fixes the Polyakov loop to $P=\pm I$ we effectively reduced the dynamics to a three-dimensional adjoint Higgs system.
We explicitly exhibited continuous paths in the $(\beta_{\rm s},\beta_{\rm H})$ plane connecting the ``deconfined symmetric'' and ``deconfined Higgs'' regimes without encountering any phase transition, thus demonstrating that the spatial dynamics alone cannot distinguish the two phases.
\end{enumerate}
These analyses give a self-consistent picture that the Higgs and deconfined regimes are not sharply separated once the temporal center symmetry is broken, although the confined phase remains distinct.

\subsection{Future directions}

\subsubsection{Determination of the phase diagram}

In the upper-left region of Figure \ref{fig:our_proposed_phase_diagram}, one finds
 the deconfinement/Higgs transition line.
We expect it to terminate at an endpoint inside the bulk region, beyond which no phase transition occurs.
According to the analysis in Section \ref{sec:global_symmetry_analysis}, this transition is expected not to be governed by Landau’s criterion.
Hence, in order to determine the endpoint, we need the analysis including full dynamics.

Lattice simulation is a promising approach for investigating the dynamics of quantum field theories.
From the viewpoint of the lattice model, the presence of the deconfinement/Higgs phase transition line is not well-established.
If it exists, it is important to find the position of the endpoint.
It is also an interesting problem to investigate the order of 
the deconfinement/Higgs phase transition.
A simple scenario is that the transition line is of first order and the endpoint is of second order.
They must ultimately be settled by further numerical investigations. 
Preliminary Hybrid Monte Carlo results are presented in Appendix~\ref{sec:Monte_Carlo_analysis}.

Another possible direction for future work is to compare the phase diagram of the lattice model at zero temperature with that at finite temperature.
There are several studies on the $\SU(2)$ adjoint Higgs lattice model at zero temperature \cite{Lang:1981qg,Brower:1982yn,Baier:1986ni,Baier:1988sc,Shibata:2023hfy,Ikeda:2023lwk}.
The phase diagrams presented in those studies are qualitatively quite different from the finite-temperature phase diagram.
It may be an interesting direction to investigate the origin of this difference.

\subsubsection{Implication for particle phenomenology and cosmology}

The existence and order of phase transitions are also important topics in particle phenomenology and cosmology, as they are closely related to phenomena such as the formation of solitons --- including monopoles, cosmic strings, and domain walls --- and bubble nucleation. Many of the discussions in these fields are based on perturbative calculations.

However, nonperturbative analyses are necessary for determining the phase structure.
A well-known example is the electroweak phase transition: Monte Carlo simulations have shown nonperturbatively that the transition is of first order for a light Higgs mass and becomes a crossover for a heavy Higgs mass \cite{Kajantie:1996mn, Kajantie:1995kf, Rummukainen:1998as,Karsch:1996yh,Gurtler:1997hr,Csikor:1998eu}.\footnote{Since the Higgs mass is now known to be 125 GeV, the electroweak transition is a crossover.}
The $\SU(2)$ adjoint Higgs model studied in this paper is in a similar situation --- 
the system may exhibit the crossover between the Higgs phase and the deconfined phase.

Progress in this direction is expected to advance the understanding of the phase structure of theories beyond the Standard Model, including GUTs.
To achieve this, it is necessary to generalize the gauge group and introduce matter fields.\footnote{For example, in the case of the $\SU(5)$ GUT, matter fields that explicitly break the center symmetry are introduced.
Hence, the  phase distinction by the center symmetry as in Section~\ref{sec:global_symmetry_analysis} is no longer valid in a strict sense.}
Since many aspects remain unexplored, further analyses are required.
In particular, GUT phase transitions are deeply related to the monopole problem in cosmology. This line of investigation may provide new insights into this long-standing issue.

\acknowledgments
The authors appreciate useful discussions with Sinya Aoki, Kohei Fujikura, Tomoya Hayata, Ryusuke Jinno, Masakiyo Kitazawa, Shinji Takeda, and Yuya Tanizaki.

The work of Y.H. was partially supported by Japan Society for the Promotion of Science (JSPS) KAKENHI Research Fellowship for Young Scientists Grant Number 23KJ1161.
The work of M.K. was partially supported by JSPS KAKENHI Grant Number 23K22491 and 25K23381.
The work of H.W. was partially supported by JSPS KAKENHI Grant Number 21J13014, 23K22489, and 24K00630.

\appendix

\section{Higgs phase in $\SU(N)$ gauge theory at zero-temperature}
\label{sec:Higgs_phase_in_SU(N)}
In this section, we consider the Higgs phase of the $\SU(N)$ adjoint Higgs model at zero temperature.\footnote{
There are works for the phase diagram of $\SU(3)$ adjoint Higgs model \cite{Gupta:1983zv,Kikugawa:1985ex}.
}
The generalization is straightforward.
In this case, the center symmetry is $\mathbb{Z}_N^{[1]}$.
Here we assume that the potential is the same as (\ref{eq:potential}). Then the Higgs pattern is
\begin{align}
\SU(N)
\xrightarrow{\rm Higgs}
\begin{cases}
\frac{\SU(k+1)
\times 
\SU(k)
\times 
\U(1)}{\mathbb{Z}_{(k+1)k}}
& \text{if $N=2k+1$,}\\
\frac{
\SU(k)
\times
\SU(k)
\times 
\U(1)}{\mathbb{Z}_{k}}
& \text{if $N=2k$.}
\end{cases}
\end{align}
In the $N=5$ case, it is
\begin{align}
\SU(5)
\xrightarrow{\rm Higgs}
\frac{
\SU(3)
\times
\SU(2)
\times
\U(1)
}
{
\mathbb{Z}_6
}.
\end{align}
This pattern is exactly the same as the $\SU(5)$ GUT.

From the above discussion,
in the limit $m^2 \to -\infty$, the center symmetry is enhanced as
\begin{align}
\mathbb{Z}_N^{[1]}
\xrightarrow{\rm enhancement}
\begin{cases}
\U(1)^{[1]}
& \text{if $N=2k+1$,}\\
\U(1)^{[1]}\times \mathbb{Z}_k^{[1]} 
& \text{if $N=2k$.}
\end{cases}
\end{align}
Due to the 't Hooft anomaly, we expect the following symmetry breaking:
\begin{align}
\begin{cases}
\U(1)^{[1]}
\xrightarrow{\rm SSB}
\mathbb{Z}_{k(k+1)}^{[1]}
\cong
\mathbb{Z}_{k}^{[1]}
\times
\mathbb{Z}_{k+1}^{[1]}
& \text{if $N=2k+1$,}
\\
\U(1)^{[1]}\times \mathbb{Z}_k^{[1]} 
\xrightarrow{\rm SSB}
\mathbb{Z}_k^{[1]} \times \mathbb{Z}_k^{[1]} 
& \text{if $N=2k$.} 
\end{cases}
\end{align}
In such a case, there appears a ``photon region," which we identify as the Higgs phase.

For $N=2k+1$, one finds
\begin{align}
\lim_{|\gamma|\to\infty}
\langle
W(\gamma)^{k(k+1)}
\rangle_{\rm symmetric}=0,\\
\lim_{|\gamma|\to\infty}
\langle
W(\gamma)^{k(k+1)}
\rangle_{\rm Higgs}\neq 0.
\end{align}
The first line is obvious because $W(\gamma)^{k(k+1)}$ has a $\mathbb{Z}_N^{[1]}$ charge.
The second line is subtle.
$W(\gamma)^{k(k+1)}$ does not carry a nontrivial charge of the IR symmetry $\mathbb{Z}_{k(k+1)}^{[1]}$:
\begin{align}
W(\gamma)^{k(k+1)}
\mapsto 
W(\gamma)^{k(k+1)},
\end{align}
because the $\mathbb{Z}_{k(k+1)}^{[1]}$ symmetry acts on the Wilson loop as $W(\gamma)
\mapsto 
e^{\frac{2\pi i n}{k(k+1)}}
W(\gamma)$ for $n \in \mathbb{Z}_{k(k+1)}$.
This implies that $
\langle
W(\gamma)^{k(k+1)}
\rangle_{\rm Higgs}\neq 0$ as $|\gamma|\to \infty$.
For $N=2k$, a similar relation holds:
\begin{align}
\lim_{|\gamma|\to\infty}
\langle
W(\gamma)^{k}
\rangle_{\rm symmetric}=0,\\
\lim_{|\gamma|\to\infty}
\langle
W(\gamma)^{k}
\rangle_{\rm Higgs}\neq 0.
\end{align}

Hence, the Wilson loop of the $k$-th power or of the $k(k+1)$-th power is the order parameter for the Higgs-confinement phase transition in the $\SU(N)$ adjoint Higgs model.

\section{Comments on phase classification in a previous study}
\label{sec:Nishimura-Ogilvie}

Nishimura and Ogilvie \cite{Nishimura:2011md} investigated the phase structure of the $\SU(2)$ adjoint gauge-Higgs model on $\mathbb{R}^3 \times S^{1}_\beta$.
The analysis is performed at the small circumference $\beta$, with the double-trace deformation.
\begin{align}
    S \longrightarrow S' = S + \gamma \int \mathrm{d}^3\vec{x}~|\tr(P)(\vec{x})|^2
\end{align}
This double-trace deformation is introduced to keep the confinement\footnote{
This deformation potential favors $P= \im \sigma^3$.
This point $P= \im \sigma^3$ is often regarded as the center symmetric point, and one could suppose that the deformation parameter $\gamma$ would be analogous to the inverse temperature.
This is true for the deformed pure Yang-Mills theory \cite{Unsal:2008ch}.
In the Polyakov gauge that diagonalizes the Polyakov loop $P = \operatorname{diag}(\rme^{\im \theta},\rme^{-\im \theta})$, the permutation $\rme^{\im \theta} \mapsto \rme^{-\im \theta}$ is a residual gauge redundancy.
Therefore, the center transformation $P= \im \sigma^3 \mapsto - \im \sigma^3$ can be undone via the permutation redundancy.
However, the situation becomes different in the adjoint Higgs model. 
In the unitary gauge $\phi = v \sigma^3$, the Polyakov loop is diagonalized in the infrared, but the permutation is no longer a gauge redundancy.
The point $P= \im \sigma^3$ (in the unitary gauge) is no longer a center-symmetric point.
Hence, the analogy between the deformation parameter and the inverse temperature is unreliable in the Higgs regime. } at small $S^1$.
The phase diagram on $(\gamma,m^2)$ plane is then studied.
In this Appendix, we revisit their phase classification.

They proposed four distinct phases in the phase diagram: the confined phase, the deconfined phase, the $(\mathbb{Z}_2\times\mathbb{Z}_2)$-broken Higgs phase, and the ``mixed confined'' phase.
This classification is based on the two 0-form global symmetries,  $ ( \mathbb{Z}_{2}^{[0]})^{\rm 3d}_{\rm center} \times ( \mathbb{Z}_{2}^{[0]} )^{\rm 3d}_{\rm Higgs}$, and the proposed order parameters are $\tr (P) $, $ \tr (P\phi) $, and $\tr (P^{2}\phi)$.

\begin{itemize}
    \item \textbf{Confined phase} (large $m^2$, large $\gamma$)

    When $m^2$ is large and $\gamma$ is large, both $ ( \mathbb{Z}_{2}^{[0]})^{\rm 3d}_{\rm center} $ and $ ( \mathbb{Z}_{2}^{[0]} )^{\rm 3d}_{\rm Higgs}$ are unbroken.
    All order parameters vanish: $\tr (P) =0$, $ \tr (P\phi) =0$, and $\tr (P^{2}\phi) =0$.
    
    \item \textbf{Deconfined phase}  (large $m^2$, small $\gamma$)

    When $m^2$ is large and $\gamma$ is small, the model becomes the pure Yang-Mills theory at high temperature.
    Hence, this regime is the deconfined phase, where $ ( \mathbb{Z}_{2}^{[0]})^{\rm 3d}_{\rm center} $ is broken while $ ( \mathbb{Z}_{2}^{[0]} )^{\rm 3d}_{\rm Higgs}$ is unbroken.
    Only $\tr (P) $ becomes nonzero: $\tr (P) \neq 0$, $ \tr (P\phi) =0$, and $\tr (P^{2}\phi) =0$.

    \item \textbf{Mixed confined phase} (large $-m^2$, large $\gamma$)

    When $-m^2$ is large and $\gamma$ is large, the system is abelianized through the Higgs mechanism, and the Polyakov loop is fixed to $P= \im \sigma^3$.
    In this phase, we have $\tr (P) = 0$, $ \tr (P\phi) \neq 0$, and $\tr (P^{2}\phi) =0$.
    Therefore, only the diagonal subgroup of $ ( \mathbb{Z}_{2}^{[0]})^{\rm 3d}_{\rm center} \times ( \mathbb{Z}_{2}^{[0]} )^{\rm 3d}_{\rm Higgs}$ survives: $ ( \mathbb{Z}_{2}^{[0]})^{\rm 3d}_{\rm center} \times ( \mathbb{Z}_{2}^{[0]} )^{\rm 3d}_{\rm Higgs} \rightarrow  ( \mathbb{Z}_{2}^{[0]} )^{\rm 3d}_{\rm diag}$.

    \item \textbf{$(\mathbb{Z}_2\times\mathbb{Z}_2)$-broken Higgs phase}   (large $-m^2$, small $\gamma$)

    When $-m^2$ is large and $\gamma$ is small, the model is in the Higgs phase at high temperature.\footnote{As shown in Figure \ref{fig:naive_phase_diagram}, the phase depends on the order of high-temperature limit $\beta \rightarrow 0$ or deep Higgs limit $m^2 \rightarrow -\infty$. Here, we suppose the deep Higgs limit is taken first. }
    Actually, in the straightforward weakly-coupled calculation for the Polyakov loop potential, one finds minima at $(P,\phi) = (\pm I,v \sigma^3)$, where only $ ( \mathbb{Z}_{2}^{[0]})^{\rm 3d}_{\rm center} $ is broken: $\tr (P) \neq 0$, $ \tr (P\phi) =0$, and $\tr (P^{2}\phi) =0$.

    However, since actual lattice calculations do not exhibit the maximal center symmetry breaking $P= \pm I$, Ref.~\cite{Nishimura:2011md} argued that the preservation of $( \mathbb{Z}_{2}^{[0]})_{\mathrm{Higgs}}$ symmetry should be regarded as a weak-coupling artifact.
    With this assumption, the $( \mathbb{Z}_{2}^{[0]})_{\mathrm{Higgs}}$ would be broken in this regime.
    Due to this reasoning, they argue that the $(\mathbb{Z}_2\times\mathbb{Z}_2)$-broken Higgs phase, where both $( \mathbb{Z}_{2}^{[0]})^{\rm 3d}_{\rm center} $ and $ ( \mathbb{Z}_{2}^{[0]} )^{\rm 3d}_{\rm Higgs}$ are broken, should appear.
    This predicts that the order parameters would take nonzero expectation values $\tr (P) \neq 0$, $ \tr (P\phi) \neq 0$, and $\tr (P^{2}\phi) \neq 0$ in this regime.
\end{itemize}

The aforementioned reasoning for the $(\mathbb{Z}_2\times\mathbb{Z}_2)$-broken Higgs phase is not conclusive. 
The interpretation of the unbroken $( \mathbb{Z}_{2}^{[0]})_{\mathrm{Higgs}}$ symmetry as a weak-coupling artifact requires more direct and substantial evidence. 
A further examination of the relevant order parameters is therefore essential to definitively determine the true pattern of symmetry breaking.
Accordingly, Appendix \ref{sec:Monte_Carlo_analysis} presents our detailed analysis of the order parameter behavior based on dedicated lattice calculations to resolve this specific issue.

\section{Monte Carlo analysis}
\label{sec:Monte_Carlo_analysis}

This appendix examines the phase structure of the lattice model using Monte Carlo simulations.
The action is given by (\ref{eq:S_latt_original}), and the conjectured phase diagram is shown in Figure~\ref{fig:PhaseDiagram_lattice}. 
As discussed in Section \ref{sec:phase_diagram_lattice_model}, the four edges of the phase diagram in terms of lattice couplings $\beta$ and $\beta_\mathrm{H}$ are analytically tractable.
Moreover, the bulk of the phase diagram was investigated numerically by a pioneering work at finite temperature \cite{Karsch:1983ps}, which gave the conjectured phase structure.
That work also addressed the question of whether the deconfined phase is separated into two distinct regions, although the available numerical evidence was not sufficient for a firm conclusion.
In this work, we discuss the deconfinement-Higgs continuity as a natural scenario and provide 
preliminary results of lattice Monte Carlos simulations of this model.

To explore the phase structure of this model, let us focus on the operators responsible for $(\mathbb{Z}_2^{[0]})^{\rm 3d}_{\rm center}$ and $(\mathbb{Z}_2^{[0]})^{\rm 3d}_{\rm Higgs}$ symmetries.
We define the Polyakov loop matrix on the lattice using the temporal link variables as 
\begin{equation}
    P(\vec{x}) = \prod_{n_{4} = 0}^{N_{\rm t}-1} U_4\left(\vec{x},n_{4} \hat{e}_{4}\right),
\end{equation}
and the trace of this matrix defines the Polyakov loop operator.
We do not utilize the expectation value of the Polyakov loop (i.e., one-point function) as an order parameter for 
$(\mathbb{Z}_2^{[0]})^{\rm 3d}_{\rm center}$ since it works correctly only in the infinite-volume limit.
Instead, we utilize the Polyakov loop correlation function (i.e., disconnected two-point function) defined as 
\begin{equation}
    C(r) 
    = 
    \langle \mathrm{tr}P(\vec{x})\,\mathrm{tr}P(\vec{y})\rangle,
    \qquad
    r = |\vec{x}-\vec{y}|.
\end{equation}
The damping of correlations at large separation implies the unbroken $(\mathbb{Z}_2^{[0]})^{\rm 3d}_{\rm center}$. In contrast, the convergence to a nonzero value is a consequence of the long-range order and can be regarded as a signal of the breaking of $(\mathbb{Z}_2^{[0]})^{\rm 3d}_{\rm center}$.

Moreover, we introduce the correlation functions of the Polyakov loop coupled to the $\SU(N)$ adjoint Higgs field~\cite{Nishimura:2011md,Karsch:1983ps} 
\begin{equation}
    C_n(r) 
    = 
    -\langle \mathrm{tr}(P^n\phi)(\vec{x})\,\mathrm{tr}(P^n\phi)(\vec{y})\rangle.
\end{equation}
For $N=2$, it is sufficient to consider $n = 1, 2$ from the symmetry perspective:
Since the operator $\tr(P^2\phi)$ has a neutral $(\mathbb{Z}_2^{[0]})^{\rm 3d}_{\rm center}$ charge, this operator is sensitive to the $(\mathbb{Z}_2^{[0]})^{\rm 3d}_{\rm Higgs}$ symmetry.
On the other hand, $\tr(P\phi)$ has a sensitivity to both $(\mathbb{Z}_2^{[0]})^{\rm 3d}_{\rm center}$ and $(\mathbb{Z}_2^{[0]})^{\rm 3d}_{\rm Higgs}$ symmetries.
If, at least, one of two $\mathbb{Z}_2$ symmetries remains unbroken, $C_1(r)$ should approach zero at $r\to \infty$.
It should be noted that the operators $\tr(P^n\phi)$ are purely imaginary.
Note also that the adjoint scalars are replaced with $\sigma^y$ in the actual simulation for the unitary gauge.\footnote{The same gauge fixing condition was imposed in \cite{Karsch:1983ps}.}

\begin{figure}[t]
    \centering
    \includegraphics[width=0.4\textwidth]{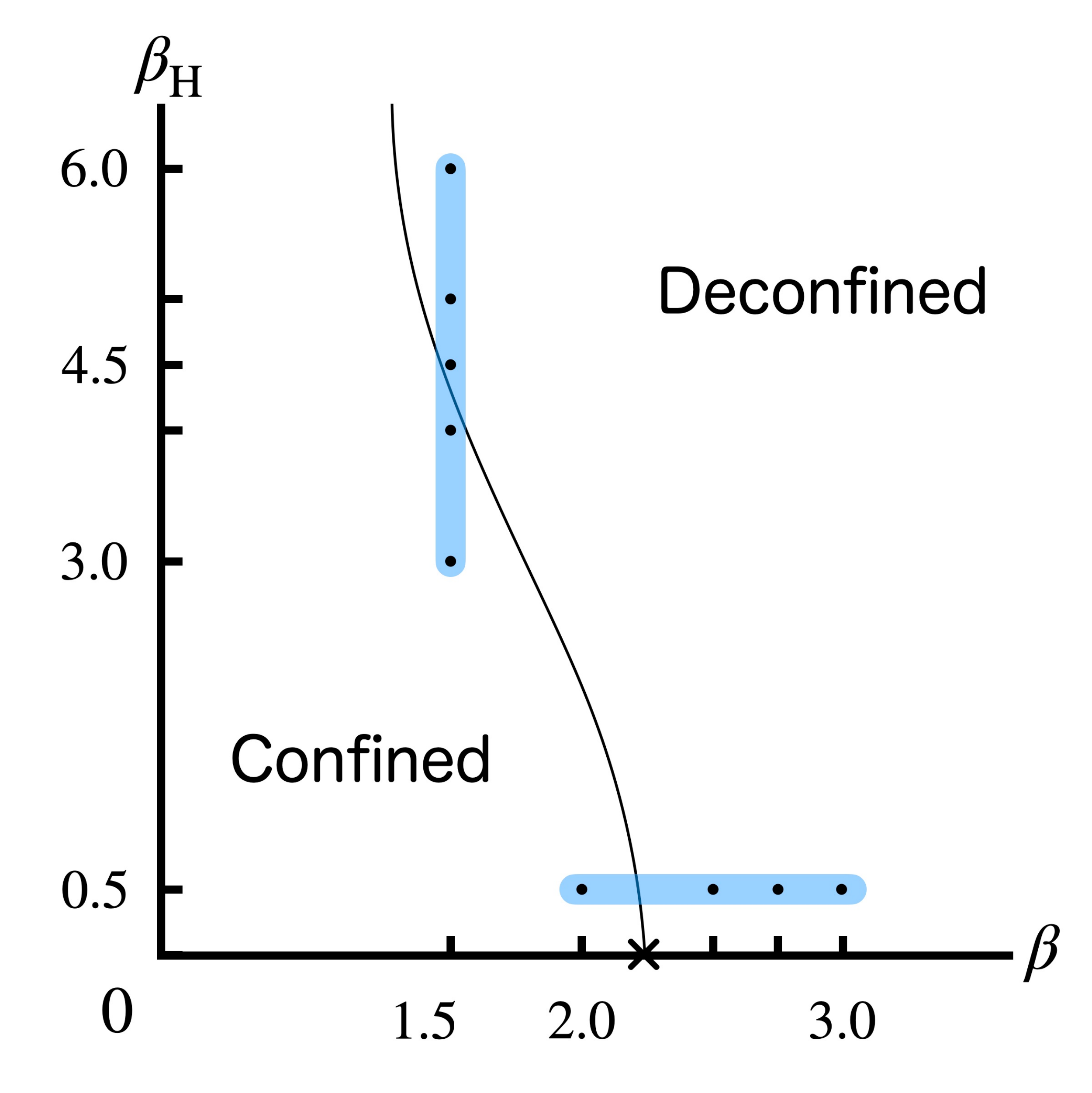}
    \caption{
    Simulation points in the schematic phase diagram of the lattice adjoint Higgs model with an $N_{\rm s}^3\times N_{\rm t} = 16^3\times 8$ lattice.
    The horizontal curve represents the confinement/deconfinement phase transition line, whose shape is imprecise.
    }
    \label{fig:PhaseDiagram_lattice_ns16nt8}
\end{figure}

\subsection{Parameter sweeps on larger lattice}

\begin{figure}[t]
    \centering
    \includegraphics[width=0.5\textwidth]{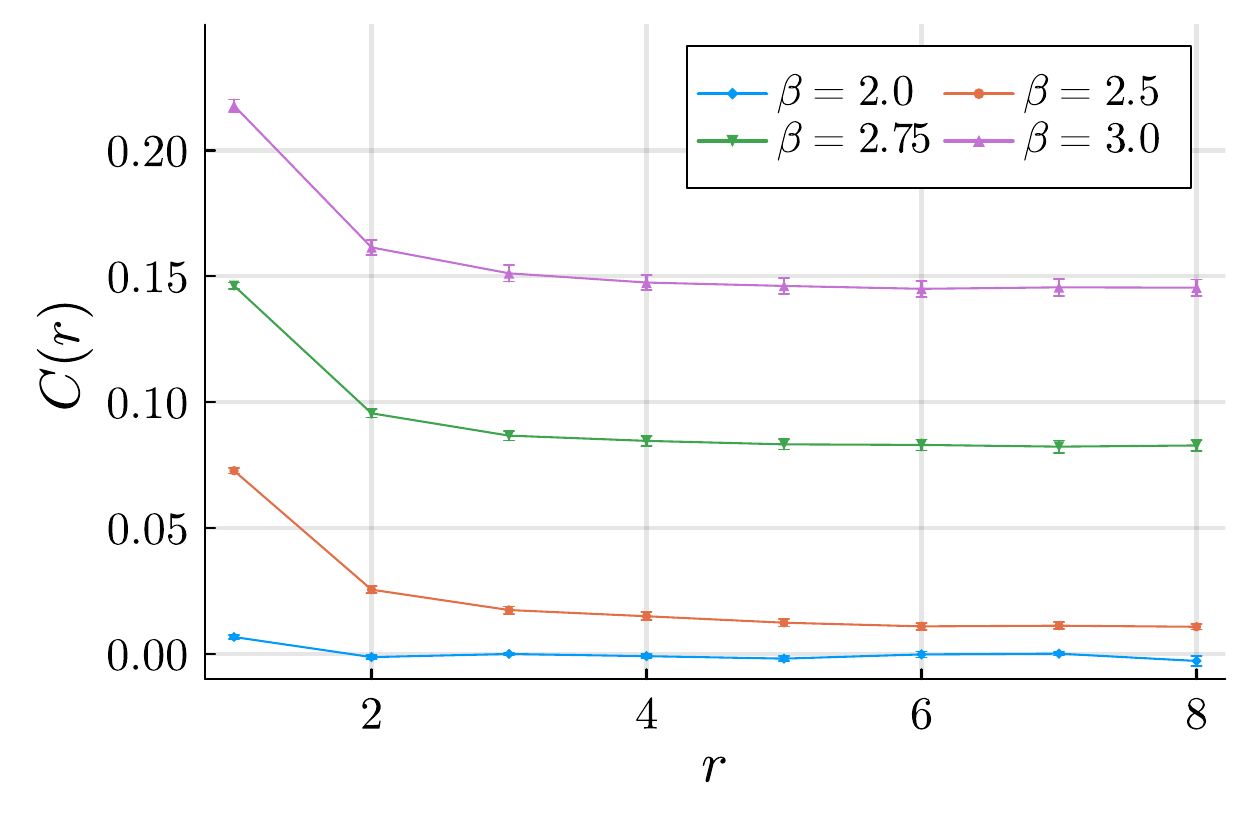}
    \caption{Plot of $C(r)$ with $\beta_\mathrm{H} = 0.5$ and $\beta = 2.0,~2.5,~2.75,~3.0$.}
    \label{fig:Corr_Pol_zoom_betaH0.5}
\end{figure}
\begin{figure}[t]
    \centering
    \includegraphics[width=0.49\textwidth]{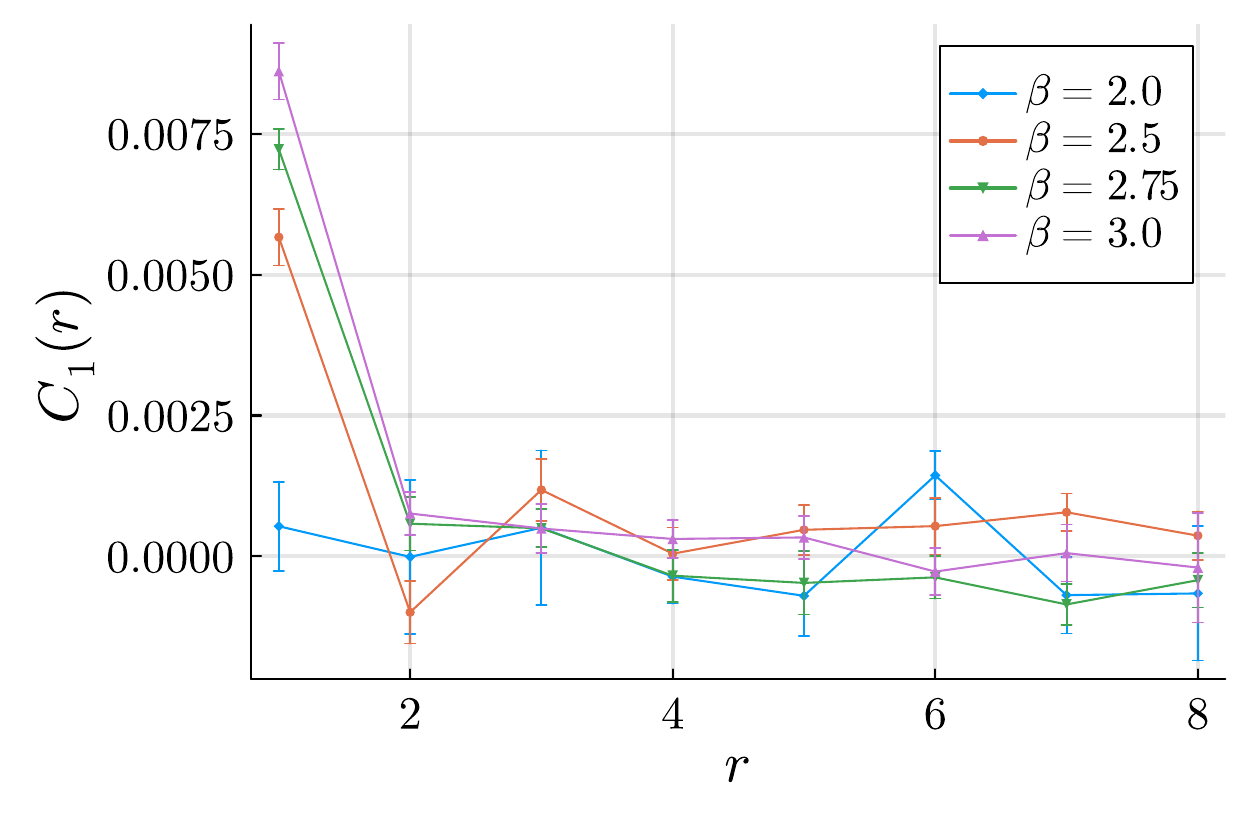}
    \includegraphics[width=0.49\textwidth]{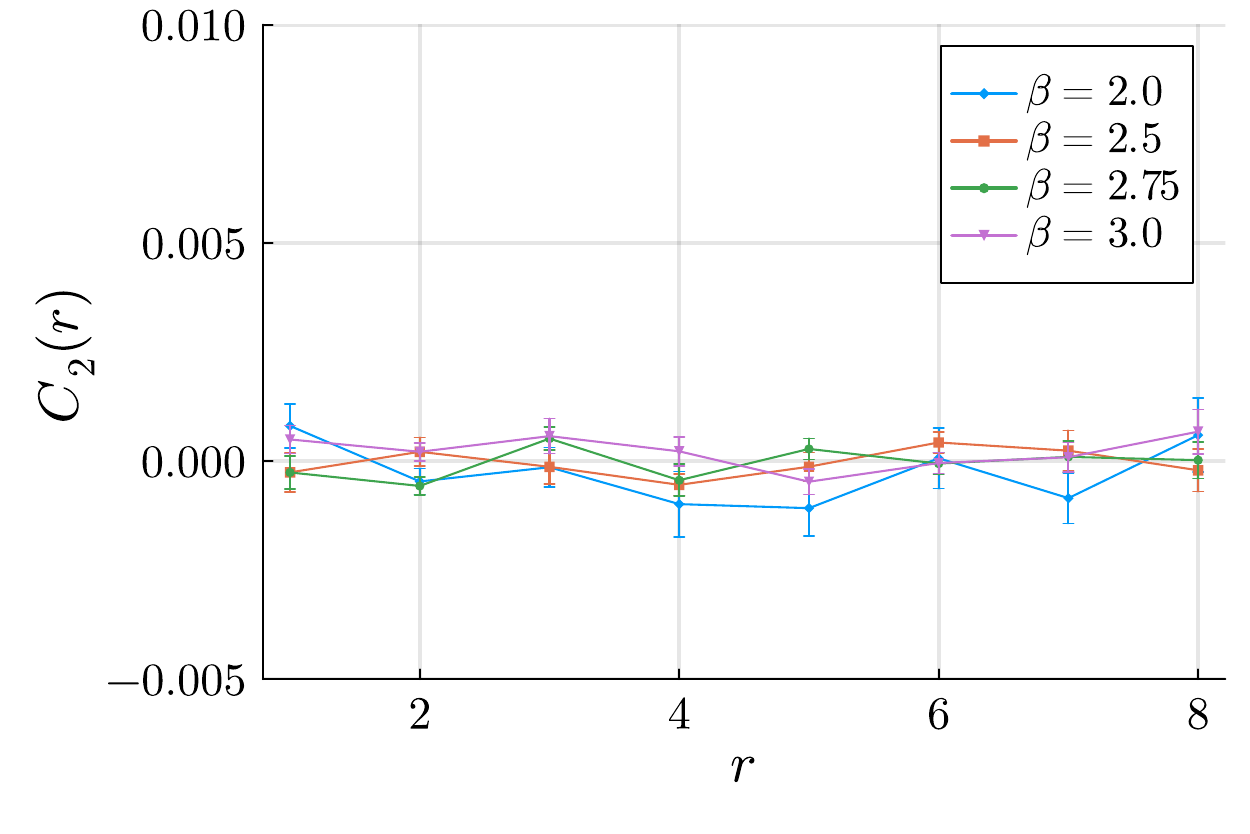}
    \caption{Plot of $C_1(r)$ (left) and $C_2(r)$ (right) with $\beta_\mathrm{H} = 0.5$ and $\beta = 2.0,~2.5,~2.75,~3.0$.}
    \label{fig:Corr_Polphi_zoom_betaH0.5}
\end{figure}

We perform two series of lattice Monte Carlo simulations with an $N_{\rm s}^3\times N_{\rm t} = 16^3\times 8$ lattice. 
More precisely, we vary the lattice couplings $\beta_\mathrm{H}$ and $\beta$ along the lines $\beta = 1.5$ and $\beta_\mathrm{H} = 0.5$, respectively. 
(See also Figure~\ref{fig:PhaseDiagram_lattice_ns16nt8}.)
\begin{table}[t]
    \centering
    \caption{Information on the gauge configurations for each lattice size and coupling.}
    \label{tab:gauge_conf}
    \begin{tabular}{|cc|cc|cc|}
        \hline
         $N_\mathrm{s}$&  $N_{\rm t}$&  $\beta$&  $\beta_\mathrm{H}$&  $N_\mathrm{conf}$& $N_\mathrm{skip}$\\\hline\hline
         16&  8&  0.5&  2.0&  400& 50\\
         &  &        &  2.5&  400& 50\\
         &  &        &  2.75&  400& 50\\
         &  &        &  3.0&  400& 50\\
         \hline
         16&  8&  1.5&  3.0&  200& 50\\
         &  &        &  4.0&  200& 50\\
         &  &        &  4.5&  200& 50\\
         &  &        &  5.0&  200& 50\\
         &  &        &  6.0&  200& 50\\
         \hline
        12& 6& 2.0& 2.8& 4500&200\\
        \hline
    \end{tabular}
\end{table}
We summarize here the generation of gauge configurations.
The numbers of gauge configurations we have generated are listed in Table~\ref{tab:gauge_conf}.
We employed the Hybrid Monte Carlo algorithm to update the link variables with the periodic boundary condition.
The configurations utilized for evaluating correlation functions are stored every $N_\mathrm{skip}$ steps after discarding the thermalization steps.
Hence, $N_\mathrm{update} = N_\mathrm{conf}\times N_\mathrm{skip}$ Monte Carlo updates are performed for each parameter.
Note that the adjoint Higgs field is eliminated from the lattice simulation since we have chosen the unitary gauge.
Note also that error bars in the analyses are estimated by jackknife analysis.

We first show the correlation functions $C(r)$ and $C_n(r)$ with fixed $\beta_\mathrm{H}$ and various $\beta$.
Figure~\ref{fig:Corr_Pol_zoom_betaH0.5} shows the Polyakov loop correlation function $C(r)$ with $\beta_\mathrm{H} = 0.5$ and $\beta = 2.0,~2.5,~2.75,~3.0$.
Since they approach nonzero values except for $\beta = 2.0$, the region $\beta \ge 2.5$ can be regarded as the deconfined (i.e., center-symmetry–broken) phase, which is consistent with the result in \cite{Karsch:1983ps}.
Moreover, Figure~\ref{fig:Corr_Polphi_zoom_betaH0.5} shows the correlation functions $C_n(r)$ with $\beta_\mathrm{H} = 0.5$ and $\beta = 2.0,~2.5,~2.75,~3.0$.
In contrast to the spontaneous breaking of $(\mathbb{Z}_2^{[0]})^{\rm 3d}_{\rm center}$, the behavior of $C_2(r)$ converging to zero up to numerical errors implies that $(\mathbb{Z}_2^{[0]})^{\rm 3d}_{\rm Higgs}$ is kept unbroken as we vary $\beta_\mathrm{H}$.

\begin{figure}[t]
    \centering
    \includegraphics[width=0.5\textwidth]{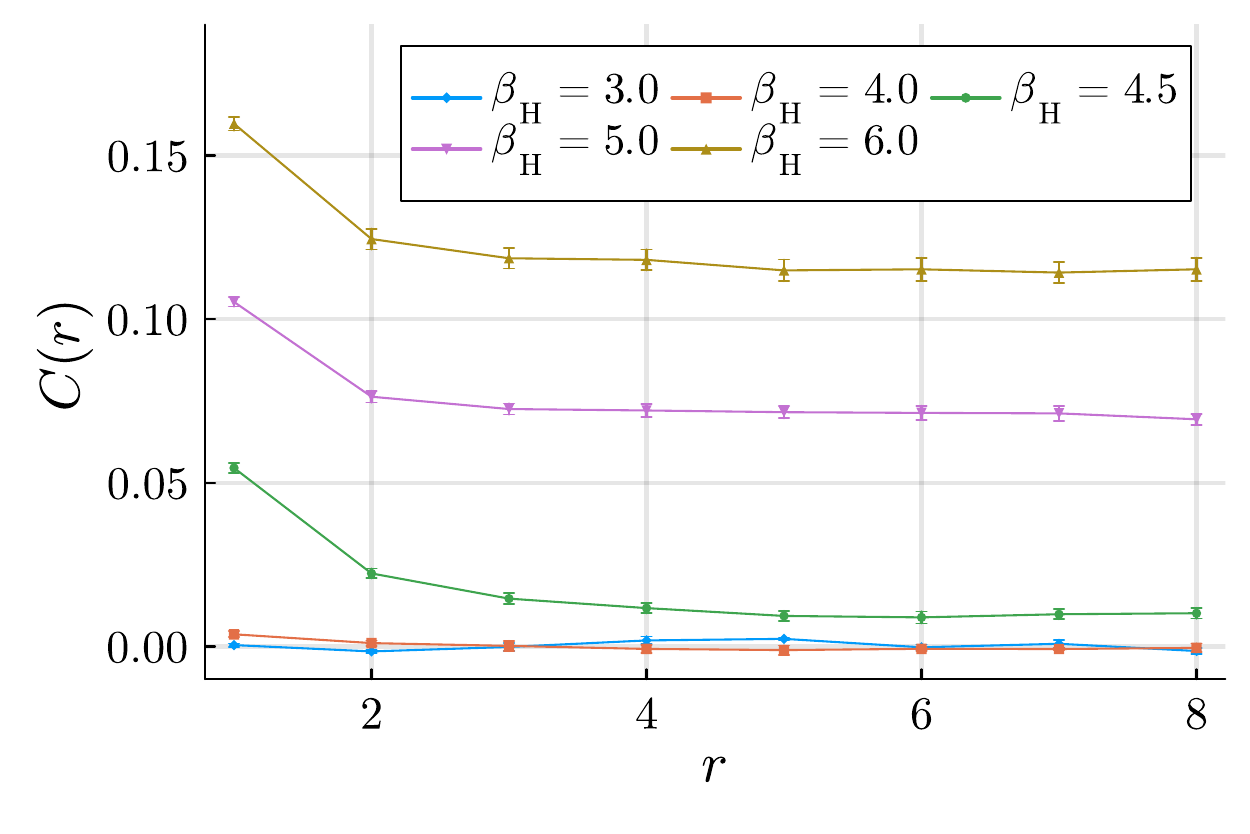}
    \caption{Plot of $C(r)$ with $\beta = 1.5$ and $\beta_\mathrm{H} = 3.0,~4.0,~4.5,~5.0,~6.0$.}
    \label{fig:Corr_Pol_zoom_beta1.5}
\end{figure}

\begin{figure}[t]
    \centering
    \includegraphics[width=0.49\textwidth]{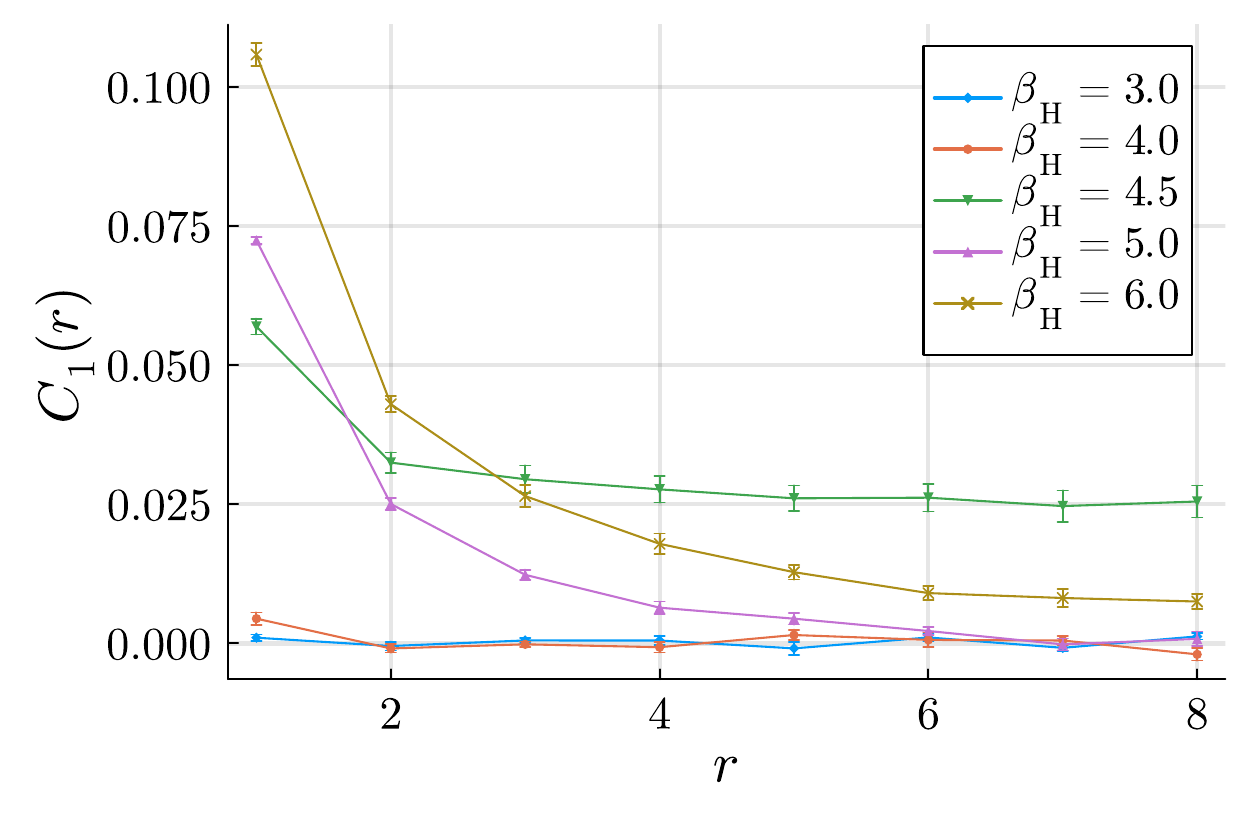}
    \includegraphics[width=0.49\textwidth]{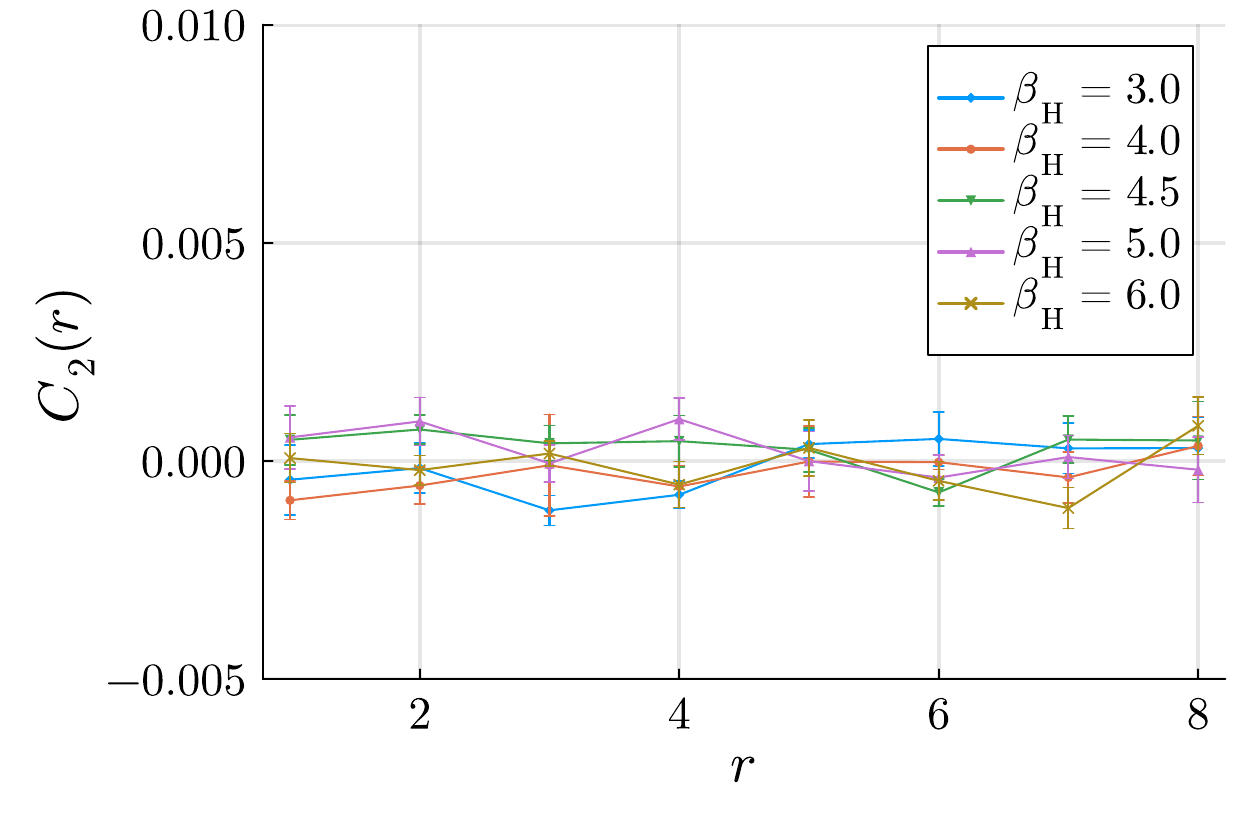}
    \caption{Plot of $C_1(r)$ (left) and $C_2(r)$ (right) with $\beta = 1.5$ and  $\beta_\mathrm{H} = 3.0,~4.0,~4.5,~5.0,~6.0$.}
    \label{fig:Corr_Polphi_zoom_beta1.5}
\end{figure}

In addition to the case with fixed $\beta_\mathrm{H}$, we perform the simulation with $\beta=1.5$ in Figures~\ref{fig:Corr_Pol_zoom_beta1.5} and \ref{fig:Corr_Polphi_zoom_beta1.5}, respectively.
From the damping of the correlators of the Polyakov loop, we confirm for $\beta = 1.5$ that the region $\beta_\mathrm{H} \ge 4.5$ is the deconfined phase as a $(\mathbb{Z}_2^{[0]})^{\rm 3d}_{\rm center}$-symmetry broken phase.
The correlation function $C_2(r)$ does not exhibit an apparent $\beta_\mathrm{H}$ dependence, even at $\beta_\mathrm{H} \gtrsim 5.0$.
Since the spontaneous breaking of $(\mathbb{Z}_2^{[0]})^{\rm 3d}_{\rm Higgs}$ symmetry is unlikely in the confined phase, this result indicates that $(\mathbb{Z}_2^{[0]})^{\rm 3d}_{\rm Higgs}$ is unbroken in the deconfined regime that we explored.


\begin{figure}[t]
    \centering
    \includegraphics[width=0.49\textwidth]{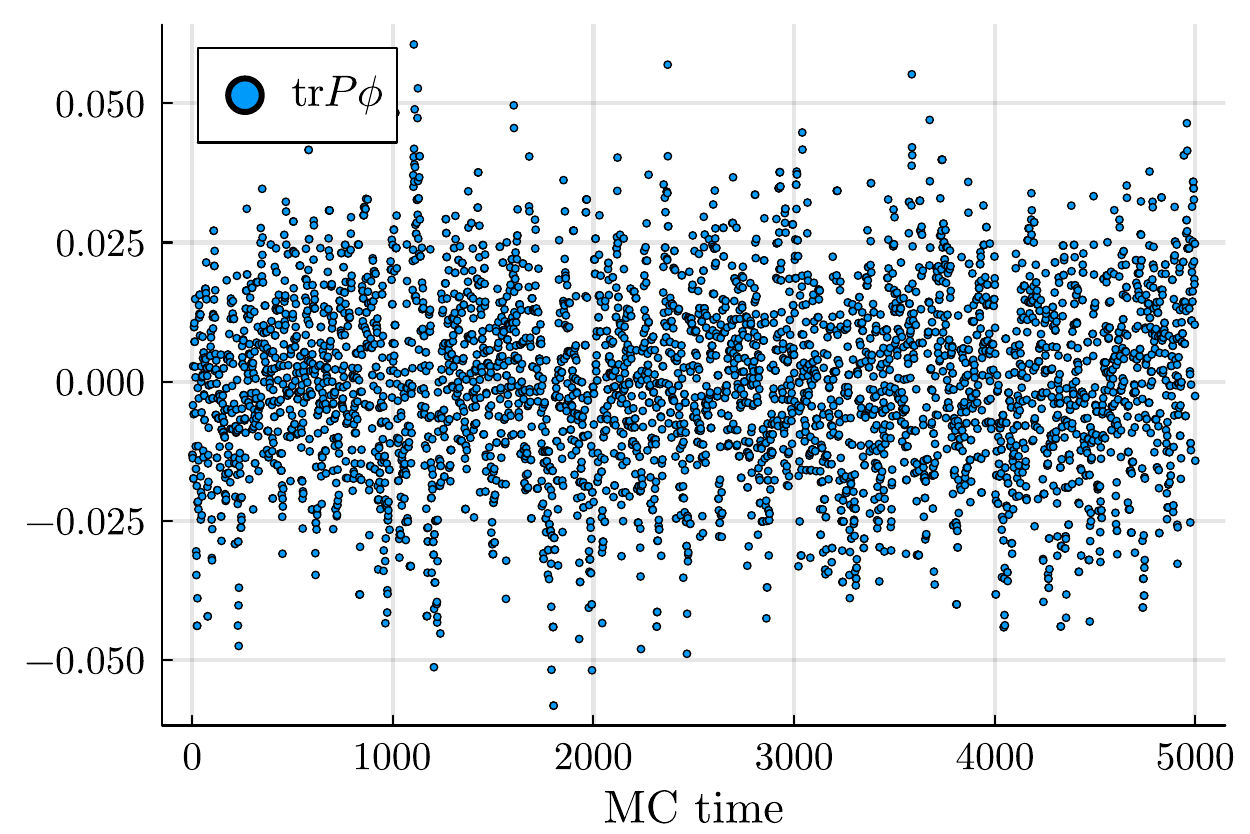}
    \includegraphics[width=0.49\textwidth]{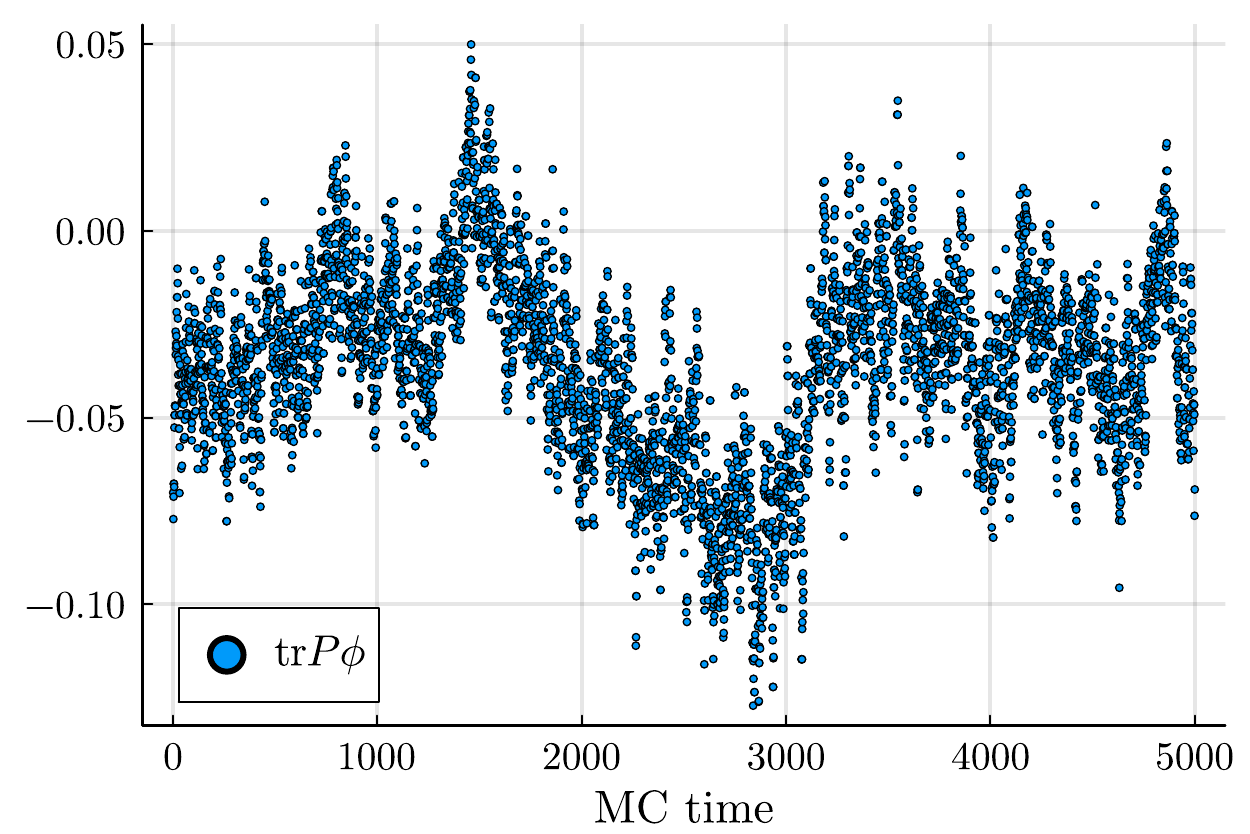}
    \caption{A part of the Monte Carlo trajectories for the imaginary part of $\tr (P\phi)$ at $\beta = 1.5,~ \beta_\mathrm{H} = 4.0$ (left) and $\beta_\mathrm{H} = 5.0$ (right).
    }
    \label{fig:MC_traj_Pphi}
\end{figure}

\begin{figure}[t]
    \centering
    \includegraphics[width=0.49\textwidth]{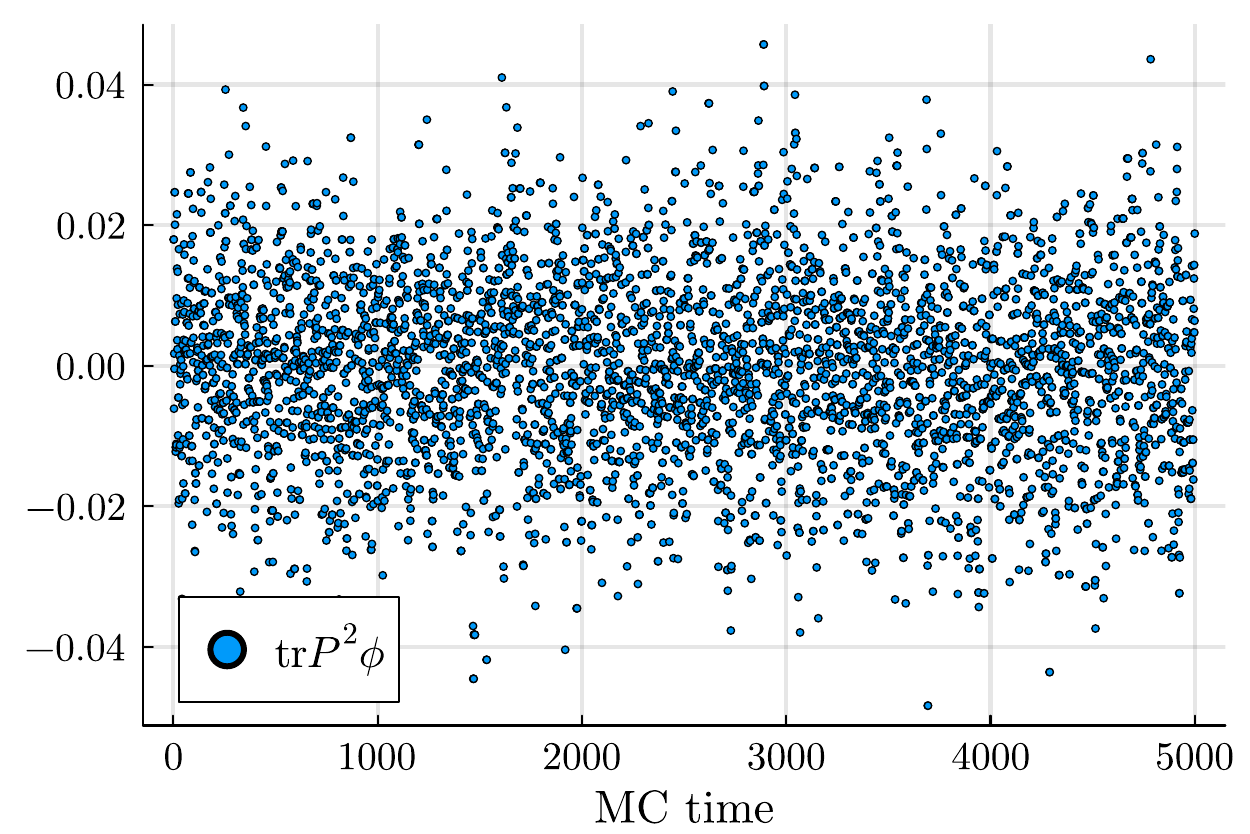}
    \includegraphics[width=0.49\textwidth]{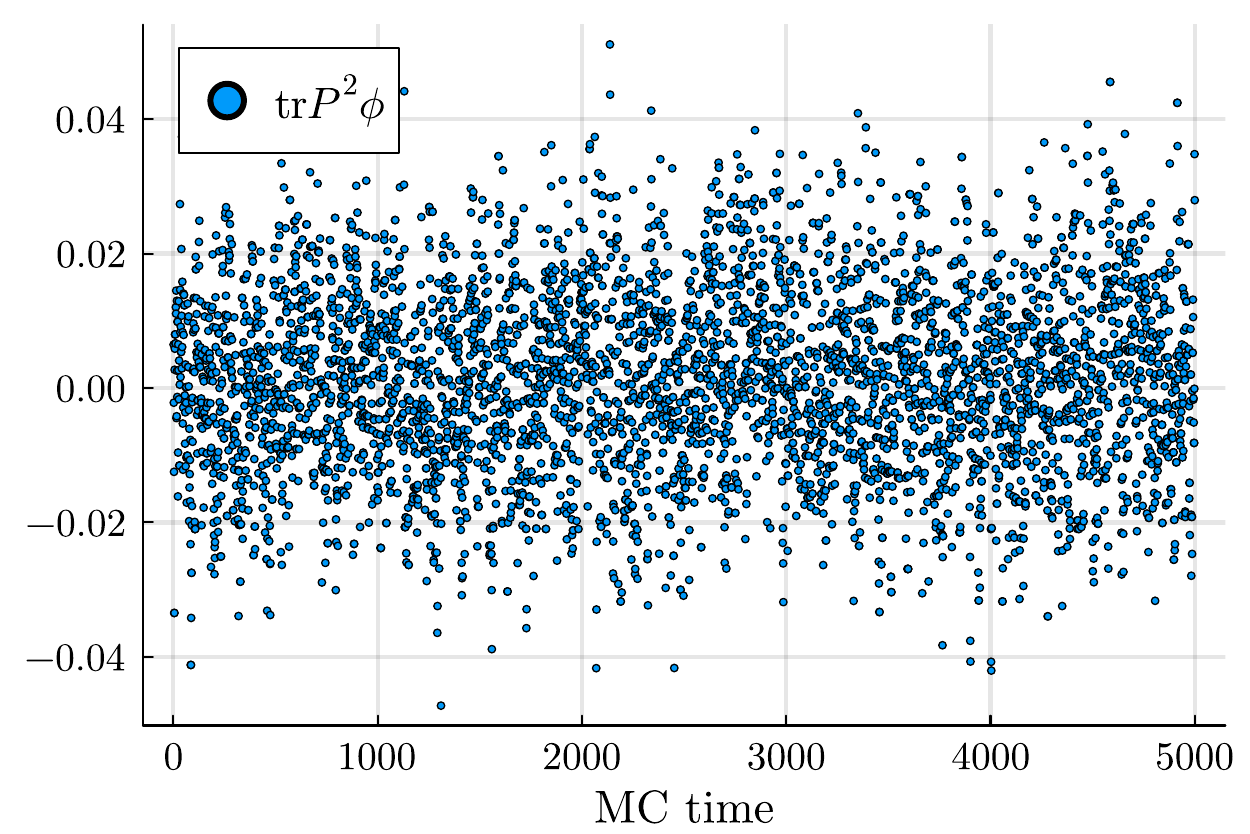}
    \caption{A part of the Monte Carlo trajectories for the imaginary part of $\tr (P^2\phi)$ at $\beta = 1.5,~ \beta_\mathrm{H} = 4.0$ (left) and $\beta_\mathrm{H} = 5.0$ (right).
    }
    \label{fig:MC_traj_P2phi}
\end{figure}

We monitor the Monte Carlo trajectory of $\tr (P\phi)$. In Figure~\ref{fig:MC_traj_Pphi}, we observe a larger fluctuation and a longer autocorrelation in the deconfined phase (right) than in the confined phase (left).
Although the plot appears to give a nonzero expectation value at first glance, using one-point functions to determine phases is quite subtle, as we mentioned before. 
This is because the transition rate among symmetry-broken vacua is small but nonzero in finite volume systems, and the expectation value will always vanish in the path-integral average\footnote{
If we were to perform a biased simulation around one of the vacua, it would violate the ergodicity that is essential for the Markov chain Monte Carlo algorithm.
}.
Furthermore, we plot the Monte Carlo trajectories for $\tr (P^2\phi)$ in Figure~\ref{fig:MC_traj_P2phi}. 
The figure shows that this operator fluctuates around zero both in the confined and deconfined phases.
This behavior is completely consistent with the rapid damping of the corresponding two-point function (on the right panel of  Figure~\ref{fig:Corr_Pol_zoom_beta1.5}).
A related analysis is conducted in Appendix~\ref{sec:eigenphase_distribution}.


A concern is that $C_1(r)$ for $\beta_\mathrm{H} = 4.5$ seems to converge to a nonzero value at large $r$.
We expect that this is due to an extended correlation length, not the $(\mathbb{Z}_2^{[0]})^{\rm 3d}_{\rm Higgs}$ breaking:
The confinement/deconfinement phase transition is of second order for the $\SU(2)$ pure lattice gauge theory in four dimensions and is expected to remain unchanged even for nonzero $\beta_\mathrm{H}$.
Moreover, the Higgs coupling $\beta_\mathrm{H}$ is presumably so close to the critical value that the correlation length is increased compared to results for other $\beta_\mathrm{H}$ values.
Although we assume that this transition is always of second order for any $\beta_\mathrm{H}$, we cannot exclude the possibility that there is a critical point that changes the order of the transition to first order.
We still expect that this issue will not drastically alter our conclusion, which supports continuity.

\subsection{Study at a single point on smaller lattice}

In this subsection, we show the result of the same analysis in a slightly larger $\beta$ region.
The Monte Carlo simulation is performed for $\beta = 2.0$ and $\beta_\mathrm{H} = 2.8$, but with a reduced lattice size of $N_{\rm s}^3\times N_{\rm t} = 12^3 \times 6$, 
which aims to enhance the tunneling among the potential minima in terms of $(\mathbb{Z}_2^{[0]})_{\rm center}^{\rm 3d}$ symmetry.
Note that this downsizing results in a trade-off, as it reduces the range over which the correlation functions can be measured.

\begin{figure}[t]
    \centering
    \includegraphics[width=0.49\textwidth]{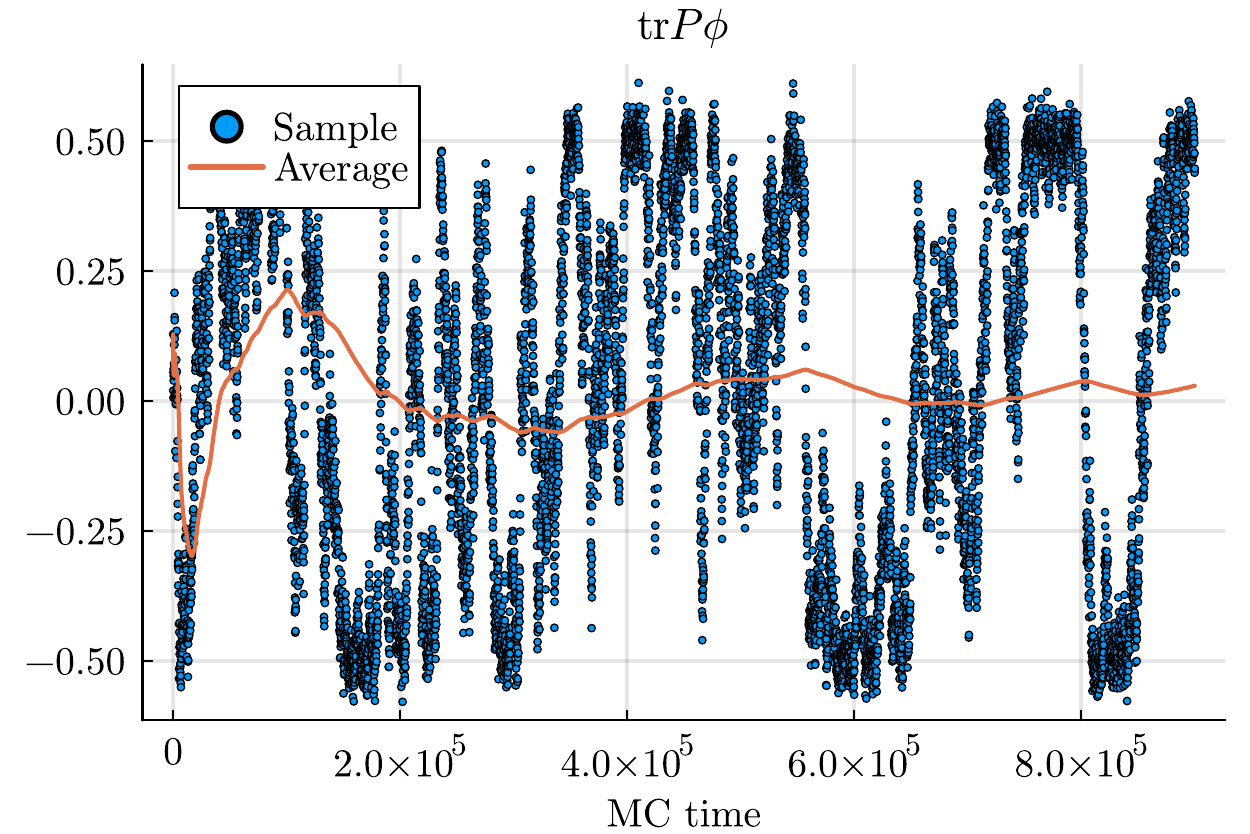}
    \includegraphics[width=0.49\textwidth]{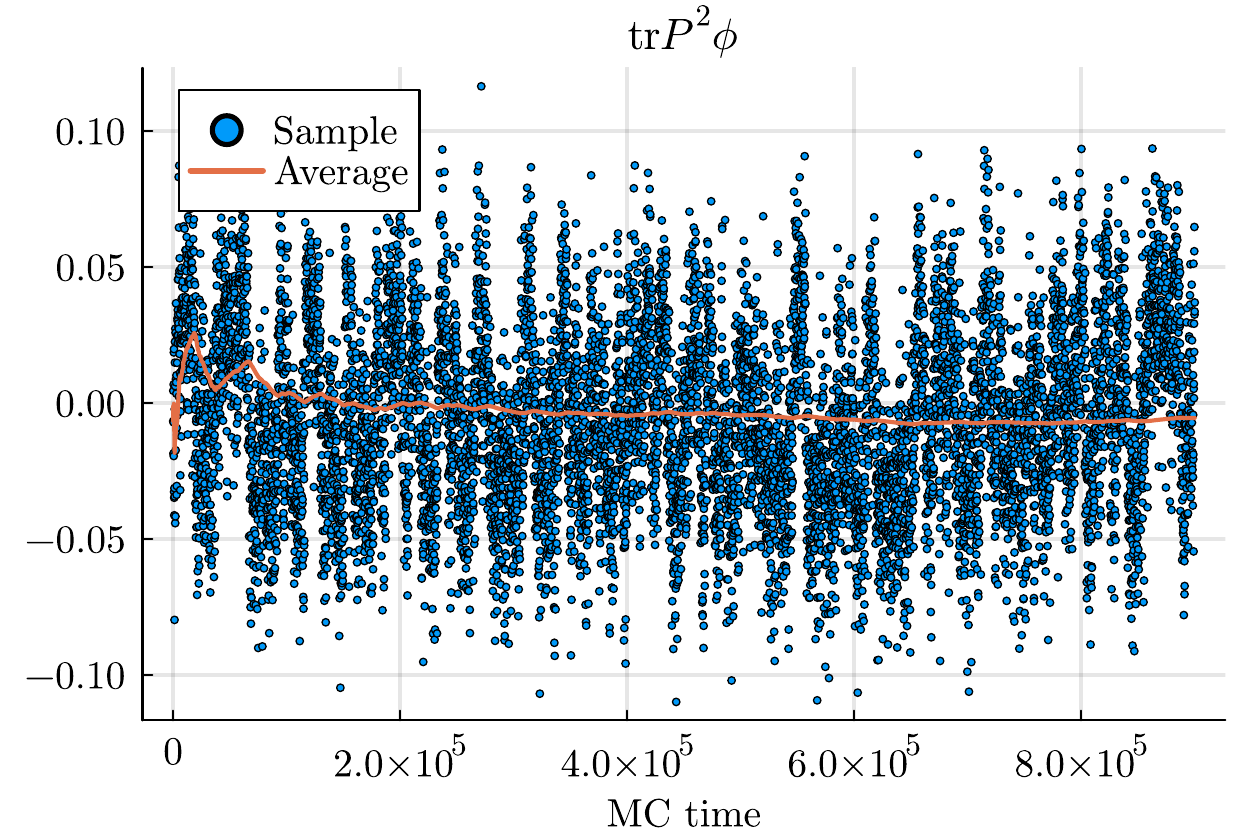}
    \caption{The Monte Carlo trajectories for the imaginary part of $\tr (P\phi)$ (left) and $\tr (P^2\phi)$  (right) at $\beta = 2.0,~ \beta_\mathrm{H} = 2.8$.
    The orange solid line represents the average at the Monte Carlo time.
    }
    \label{fig:MC_hist_traj_polphi_ns12nt6}
\end{figure}
\begin{figure}[t]
    \centering
    \includegraphics[width=0.49\textwidth]{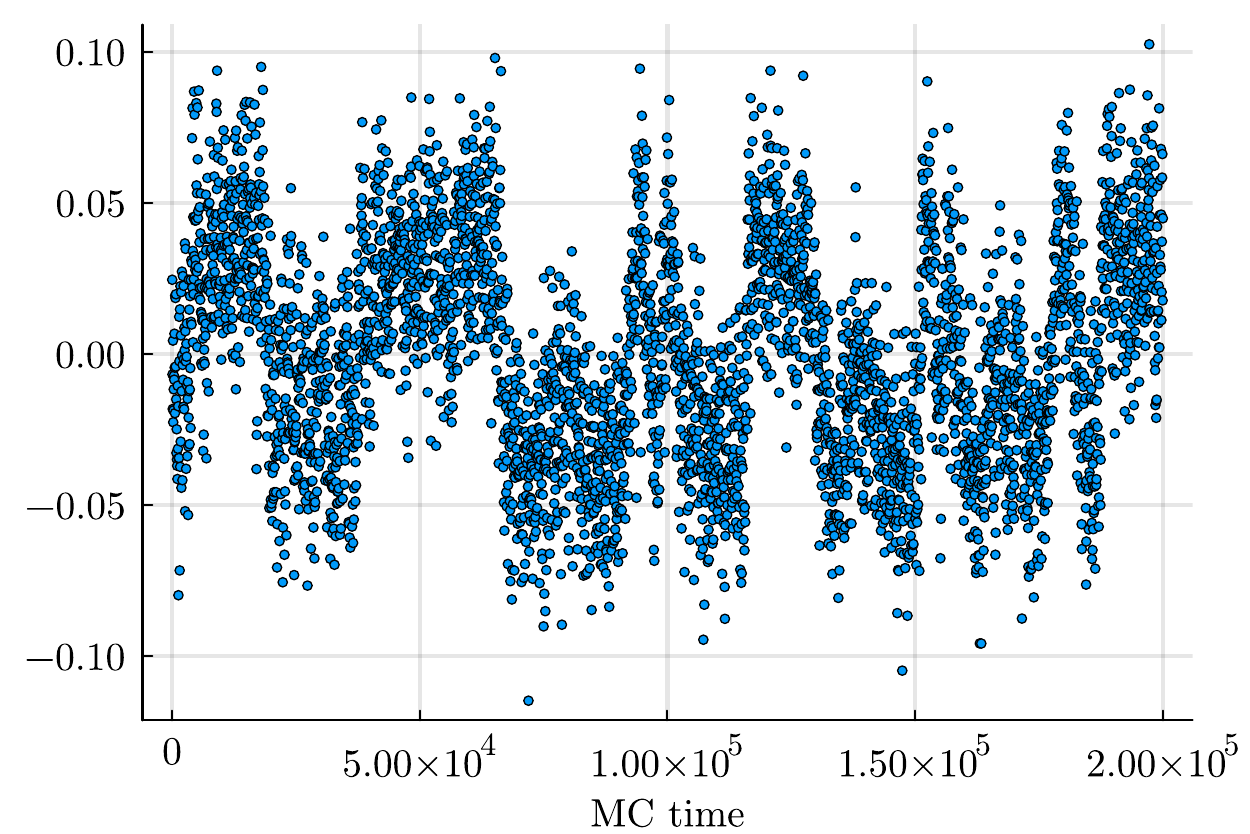}
    \caption{The Monte Carlo trajectory for the imaginary part of $\tr (P^2\phi)$ at $\beta = 2.0,~ \beta_\mathrm{H} = 2.8$, up to $200,000$ Monte Carlo steps. 
    }
    \label{fig:MC_hist_traj_pol2phi_ns12nt6_zoom}
\end{figure}

The reason we have examined this point is that the operator $\tr(P^2\phi)$ exhibits a rather different behavior from that in the previous subsection.
In Figure~\ref{fig:MC_hist_traj_polphi_ns12nt6}, we show the Monte Carlo trajectories for the imaginary part of $\tr (P \phi)$ (left) and $\tr (P^2 \phi)$ (right).
In Figure~\ref{fig:MC_hist_traj_pol2phi_ns12nt6_zoom}, we extract the first 200,000 Monte Carlo steps of the imaginary part of $\tr (P^2\phi)$ from Figure~\ref{fig:MC_hist_traj_polphi_ns12nt6}.
These plots show a long autocorrelation, i.e., the correlation in the Monte Carlo time direction: Specifically, the sampling fluctuates around certain positive and negative values with the same magnitude and frequently jumps among them along the simulation.
This apparent hysteresis indicates the existence of two minima for the effective potential.
Combining it with the observation that such a hysteresis is not seen from the previous analysis for small $\beta_\mathrm{H}$, it is tempting to connect this behavior with a phase transition induced by the spontaneous breaking of $(\mathbb{Z}_2^{[0]})^{\rm 3d}_{\rm Higgs}$ symmetry.
However, we should be careful: The orange lines represent the Monte Carlo average of corresponding operators, i.e., the one-point functions $\ev{\tr (P\phi)}$ (left) and $\ev{\tr (P^2\phi)}$ (right), and both are close to zero. 
This is compatible with the fact that spontaneous symmetry breaking does not occur in the strict sense for a finite system when one performs a simulation that maintains ergodicity.

\begin{figure}[t]
    \centering
    \includegraphics[width=0.5\textwidth]{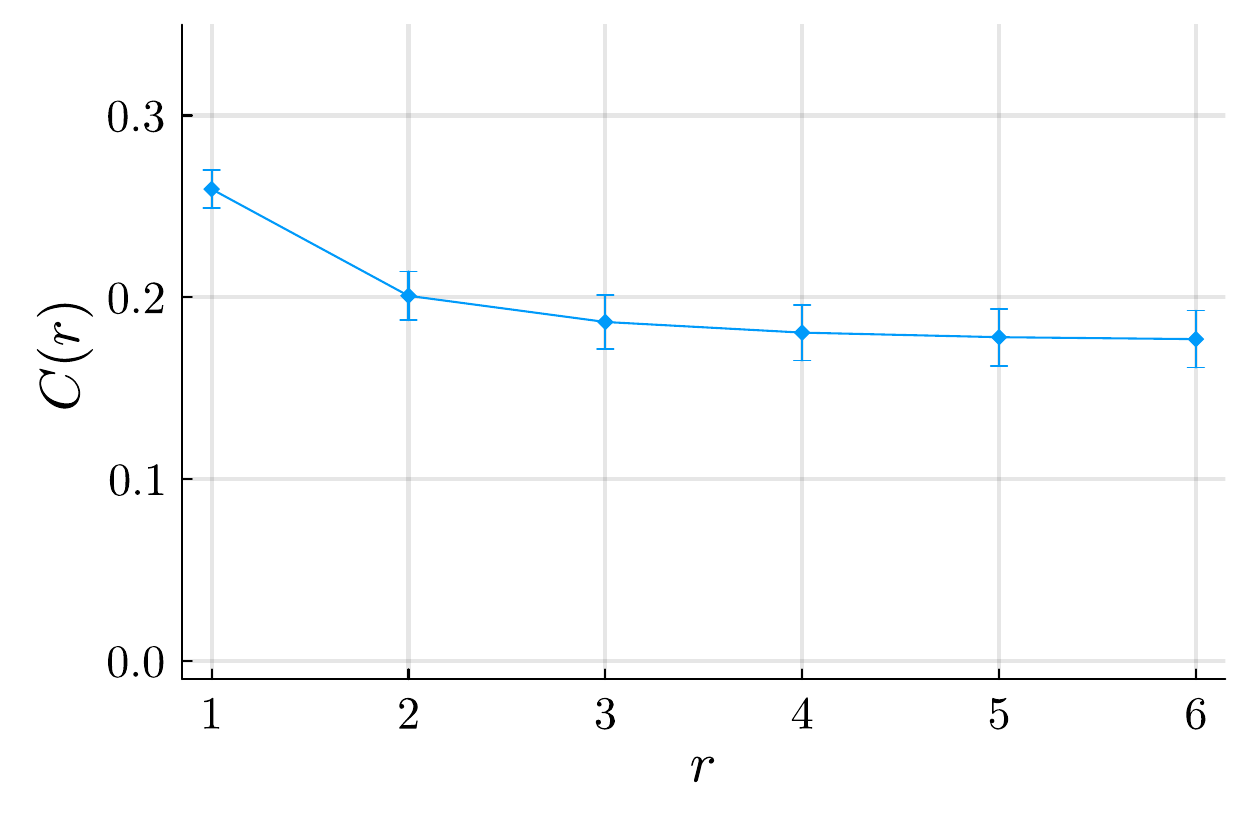}
    \caption{Plot of $C(r)$ at $\beta = 2.0$ and $\beta_\mathrm{H} = 2.8$.}
    \label{fig:Corr_Pol_zoom_ns12nt6}
\end{figure}

\begin{figure}[t]
    \centering
    \includegraphics[width=0.49\textwidth]{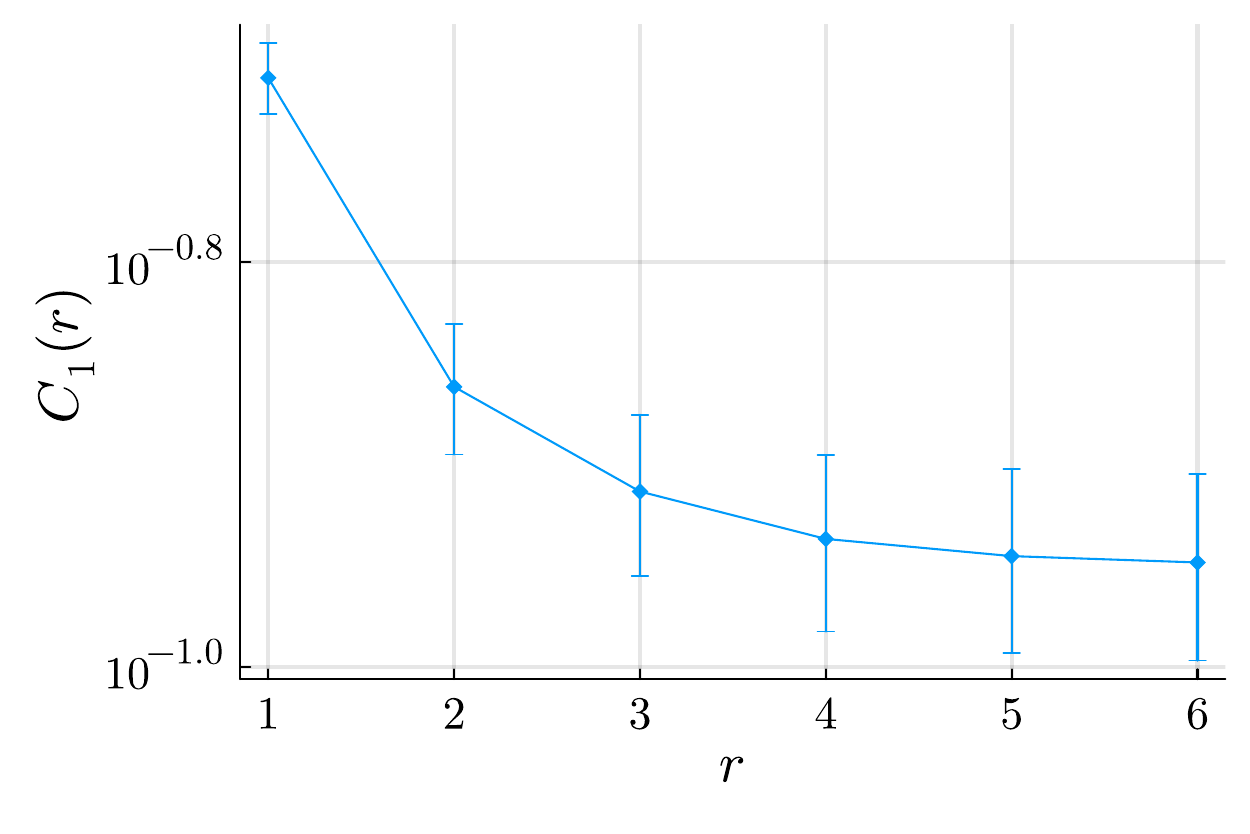}
    \includegraphics[width=0.49\textwidth]{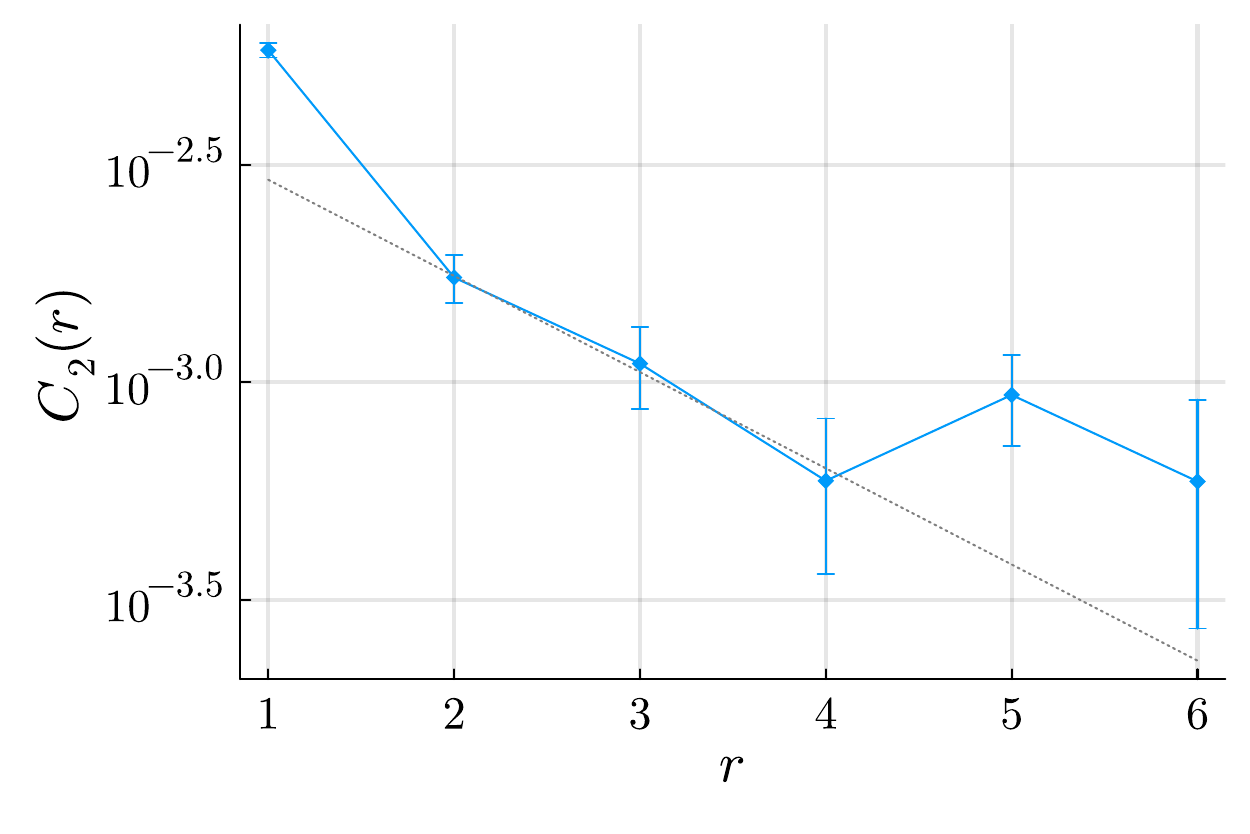}
    \caption{Plot of $C_1(r)$ (left) and $C_2(r)$ (right) at $\beta = 2.0$ and $\beta_\mathrm{H} = 2.8$.}
    \label{fig:Corr_Polphi_zoom_ns12nt6}
\end{figure}

To see if $(\mathbb{Z}_2^{[0]})^{\rm 3d}_{\rm Higgs}$ symmetry breaking occurs or not, we measure the correlation functions composed of the Polyakov loop operator and the Higgs field.
We plot $C(r)$ in Figure~\ref{fig:Corr_Pol_zoom_ns12nt6} as a function of distance $r$.
It shows convergence to a nonzero value within the spatial range we can take, and hence can be interpreted as the deconfinement at this parameter.
In Figure~\ref{fig:Corr_Polphi_zoom_ns12nt6}, we next plot $C_1(r)$ (left) and $C_2(r)$ (right) as a function of $r$, on a logarithmic scale for the vertical axes.
The dotted line on the right panel represents the fitting line with the function $A \exp(-Br)$ for $2 \le r \le 4$.
For $C_2(r)$ (right), the exponential damping for short range
is clear. 
For $r=5,6$, the result is not inconsistent with the exponential damping within the current level of statistical accuracy, although a small deviation is observed. \footnote{Due to the periodic boundary condition on the lattice, the maximal length of separation is $N_{\rm s}/2$. 
As $r$ becomes larger, the signal-to-noise ratio tends to decrease.}
The result can be regarded as suggestive evidence of the unbroken $(\mathbb{Z}_2^{[0]})^{\rm 3d}_{\rm Higgs}$ symmetry.

Before closing this section, we should emphasize that, at the current stage, it is insufficient to establish the deconfined-Higgs continuity numerically.
First, phase transitions become exact only in the infinite-volume limit, namely $N_{\rm s} \to \infty$. 
We have performed simulations for two different lattice volumes but with different lattice couplings. 
It is essential to verify that the characteristic behavior in thermal quantities around the transition point becomes increasingly pronounced as we increase $N_{\rm s}$.
Without such a finite-size scaling analysis, the present scenario regarding the deconfinement–Higgs continuity remains inconclusive.

Moreover, the consistent signals of the unbroken $(\mathbb{Z}_2^{[0]})^{\rm 3d}_{\rm Higgs}$ symmetry are obtained only in a limited region of the deconfined phase.
The hysteresis phenomenon for $\tr(P^2\phi)$ is observed in contrast with the small $\beta_{\rm H}$ region. 
A possible interpretation is that this originates from the deconfinement/Higgs transition and that the simulation point $(\beta,\beta_\mathrm{H}) = (2.0,2.8)$ is close to the transition line.
If this interpretation is indeed valid, we have indirectly captured the deconfinement/Higgs transition, which is a significant advancement toward unveiling the full structure of the phase diagram.
To further verify our proposal regarding the deconfinement-Higgs continuity, it is necessary to demonstrate that the transition is not induced by the spontaneous breaking of $(\mathbb{Z}_2^{[0]})^{\rm 3d}_{\rm Higgs}$ symmetry but by the dynamics of this model. 
This issue is left for future studies.

The result for $C_1(r)$ is also difficult to give a fair statement: 
It can be interpreted either way, in particular, at large $r$.
If $(\mathbb{Z}_2^{[0]})^{\rm 3d}_{\rm Higgs}$ is unbroken, this correlation function, which is responsible for the $(\mathbb{Z}_2^{[0]})^{\rm 3d}_{\rm center}\times(\mathbb{Z}_2^{[0]})^{\rm 3d}_{\rm Higgs}$ symmetry, should decay in the long-range limit $r\to \infty$. 
The autocorrelation is an operator-dependent observable, and in fact, the correlation along the Monte Carlo time for $\tr(P\phi)$ seems longer than that for $\tr(P^2\phi)$ as seen in Figure~\ref{fig:MC_hist_traj_polphi_ns12nt6}.
It is a more difficult task than for $C_2(r)$ to specify whether the behavior of $C_1(r)$ is due to the long autocorrelation or the spontaneous breaking of the $(\mathbb{Z}_2^{[0]})^{\rm 3d}_{\rm Higgs}$ symmetry.
If the deconfinement/Higgs transition is of second order and occurs close to the simulation point, we expect an increased correlation length that is similar to the above case at $(\beta,\beta_\mathrm{H}) = (1.5,4.5)$.
Note that a further radical scenario, in which there is another critical point on the line that alters the order of transition, is not ruled out.
Hence, the nature of the deconfinement/Higgs transition makes the situation more intricate, and the problem cannot be resolved in a straightforward manner.
We revisit these obstacles at the end of this appendix.

\begin{figure}[t]
    \begin{minipage}[b]{0.5\hsize}
        \centering
        \includegraphics[width=\textwidth]{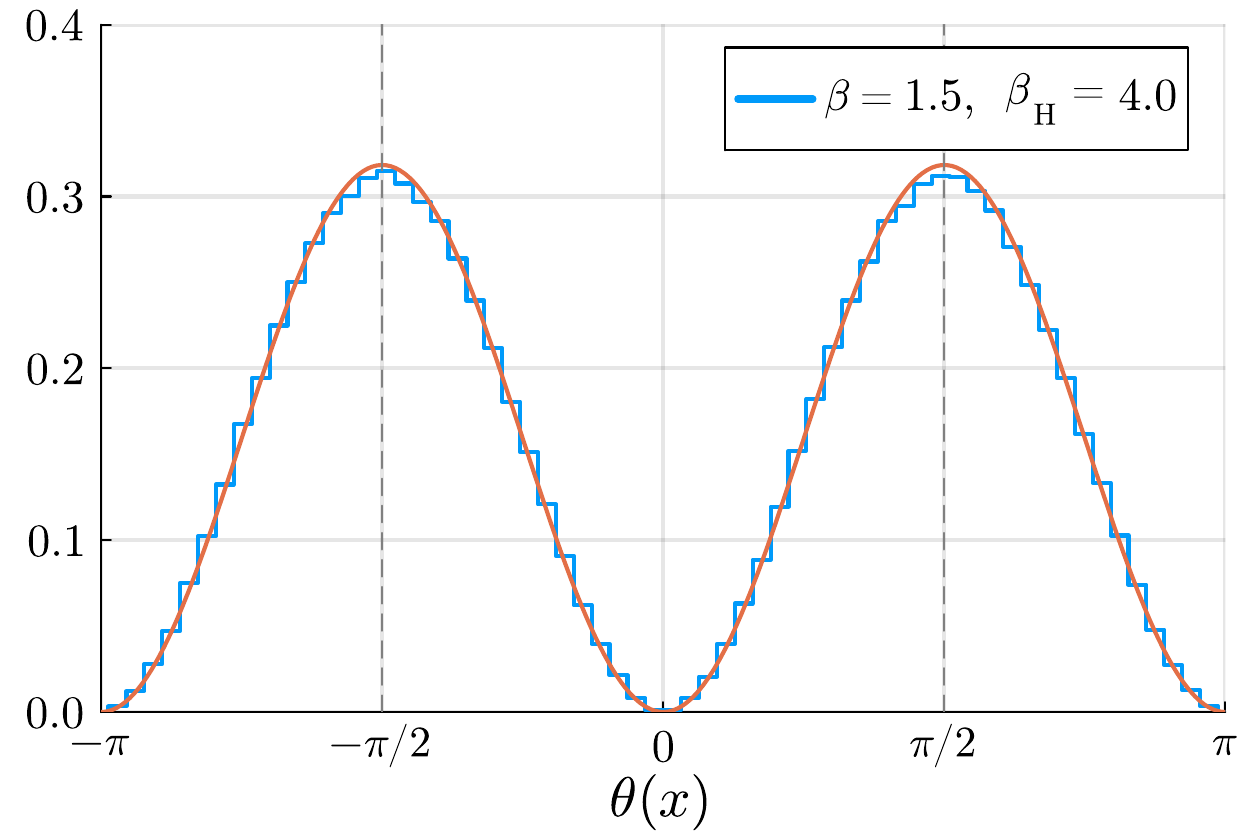}
        \subcaption{$\beta_\mathrm{H}=4.0$}
    \end{minipage}
    \begin{minipage}[b]{0.5\hsize}
        \centering
        \includegraphics[width=\textwidth]{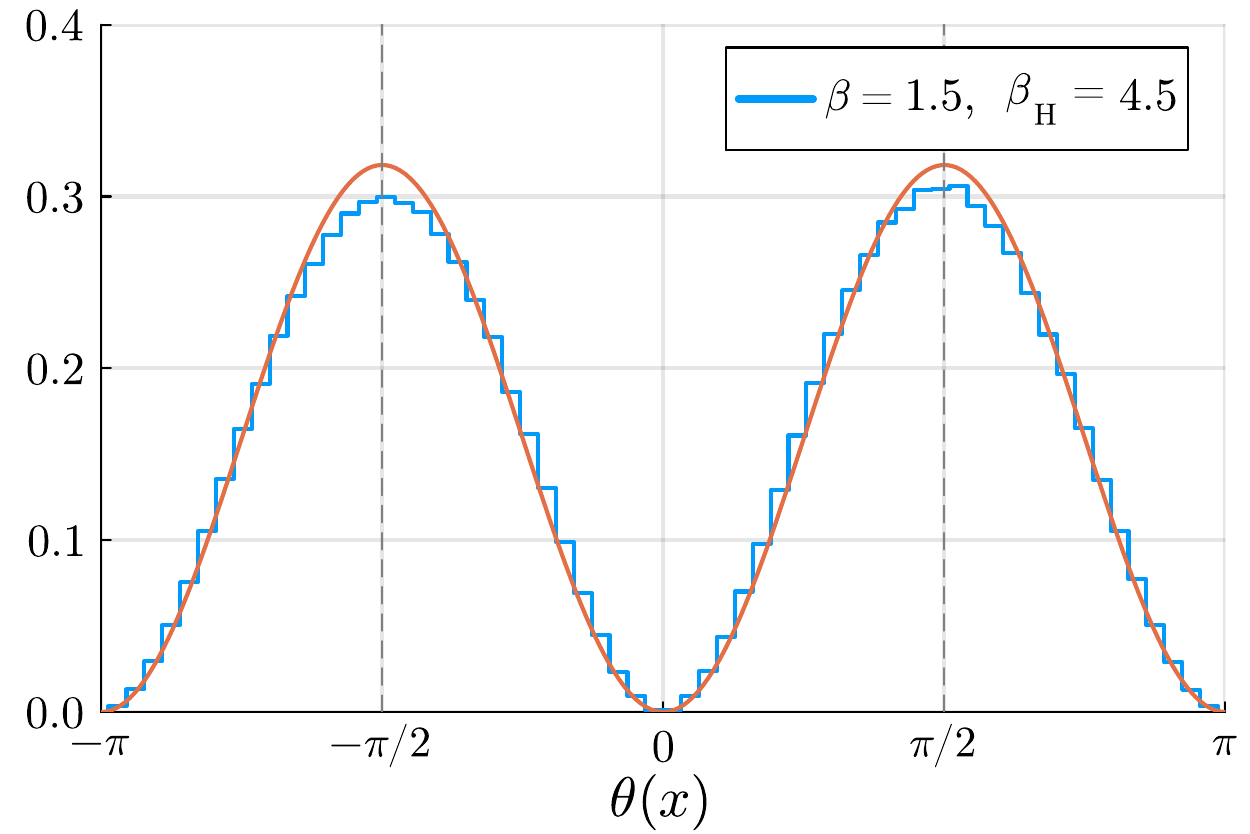}
        \subcaption{$\beta_\mathrm{H}=4.5$}
    \end{minipage}
    \begin{center}
    \begin{minipage}[b]{0.5\hsize}
        \centering
        \includegraphics[width=\textwidth]{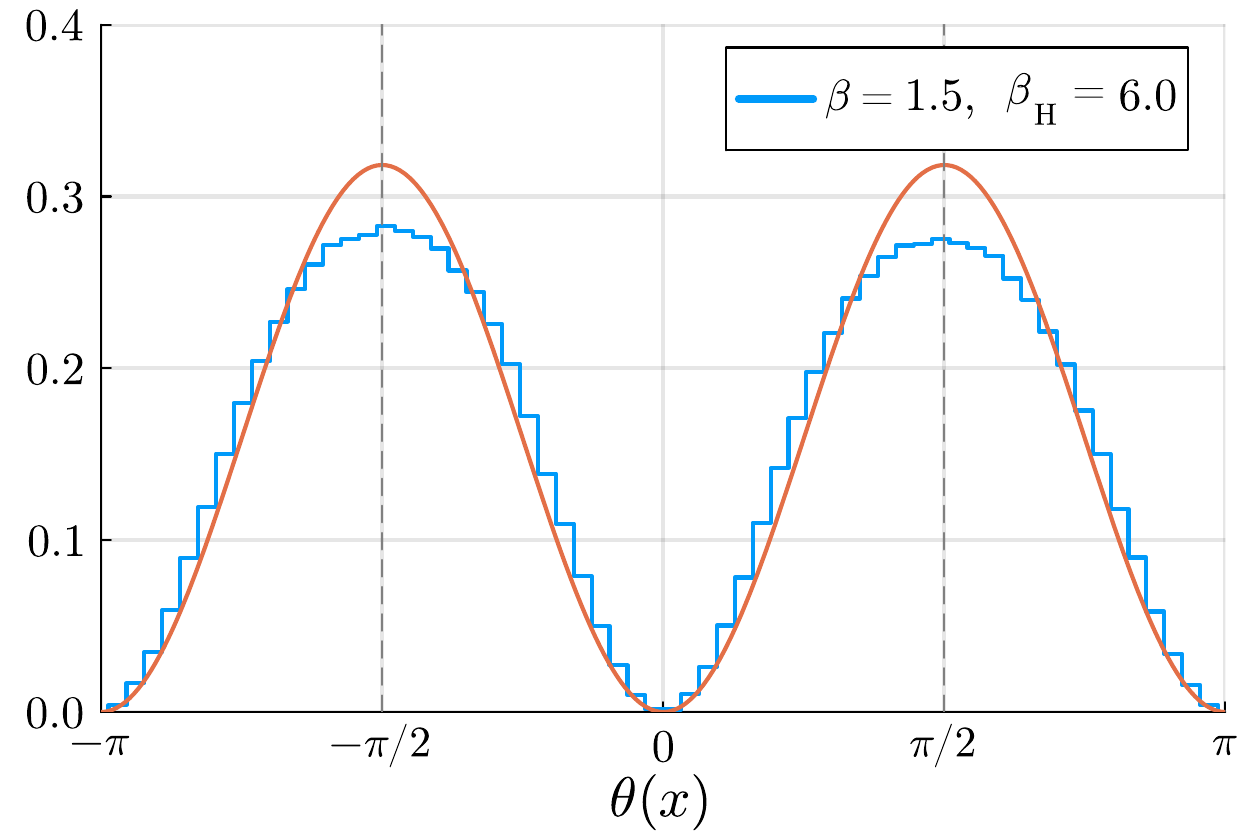}
        \subcaption{$\beta_\mathrm{H}=6.0$}
    \end{minipage}
    \end{center}
    \caption{
    Plots of the Polyakov loop eigenphase distribution at $\beta = 1.5$ and several $\beta_\mathrm{H}$.
    The orange lines represent the Haar-random distribution $\rho_\mathrm{Haar}(\theta)$.
    }
    \label{fig:eigenphase_distrib_beta1.5}
\end{figure}

\subsection{Polyakov loop eigenphase distribution}
\label{sec:eigenphase_distribution}

We here discuss the phase classification of the lattice model from a different viewpoint, namely, in terms of the Polyakov loop eigenphases.
Since the Polyakov loop matrix (before taking the trace) is an $\SU(2)$ matrix, one can diagonalize it by a certain $V \in \SU(2)$ as
\begin{equation}
    P(\vec{x})
    =
    V\diag(\ee^{\im\theta(\vec{x})},\ee^{-\im\theta(\vec{x})})V^{-1}.
\end{equation}
For the eigenphase $\theta(\vec{x})$, we can define the following distribution function 
\begin{equation}
    \rho(\theta) = \ev{\frac{1}{N_\mathrm{s}^3}\sum_{\vec{x}}\delta\big(\theta - \theta(\vec{x})\big)}.
\end{equation}
When taking the infinite-volume limit, the eigenphase distribution becomes a continuous function and may be useful to capture the phase transition~\cite{Hanada:2023krw,Hanada:2023rlk} since $\theta(\vec{x})$ has essentially the same information as the local Polyakov loop.
It is important to distinguish between $\theta(\vec{x})$ and the so-called average phase $\Theta$ of the Polyakov loop $L=\ev{\tr P(\vec{x})}_\mathrm{spatial}=\abs{L}\ee^{\im\Theta}$, as they are conceptually different despite their apparent similarity. 
The former is defined locally and therefore has rich local information, and $\rho(\theta)$ reflects the correlation among them, whereas the latter is obtained after taking the spatial average and is distributed 
equally around $\Theta = \frac{2\pi n}{N_\mathrm{c}},~(n=0,1,\cdots,N_\mathrm{c}-1)$ in the deconfined phase.

It was pointed out~\cite{Hanada:2023krw,Hanada:2023rlk,Bergner:2024waa} that the eigenphase distribution in the confined phase agrees with the $\SU(2)$ Haar-random distribution\footnote{
The analytic formulas for $\rho_\mathrm{Haar}(\theta)$ in the circular orthogonal, unitary, and symplectic ensembles (COE, CUE, CSE) are given in \cite{Nishigaki:2024phx}, where the analysis is based on the technique from random matrix theory. 
}
\begin{equation}
    \rho_\mathrm{Haar}(\theta) 
    =
    \frac{1}{2\pi}\big(1-\cos(2\theta)\big).
    \label{eq:SU(2)-Haar-random}
\end{equation}
As discussed later, our simulation also observed that the eigenphase distribution is close to \eqref{eq:SU(2)-Haar-random} in the confined phase, and deviates from it in the deconfined phase.

Figure~\ref{fig:eigenphase_distrib_beta1.5} plots the distribution of the Polyakov loop eigenphases $\rho(\theta)$ with $(N_\mathrm{s},N_{\rm t}) = (16,8)$ and fixed $\beta$ and various $\beta_\mathrm{H}$.
The distributions for $\beta_\mathrm{H} = 4.0$ agree well with the Haar-random distribution $\rho_\mathrm{Haar}(\theta)$ drawn by the orange line, and an evident discrepancy can be seen for $\beta_\mathrm{H} \ge 4.5$.
Since the sign of the Polyakov loop $\tr (P)$ is flipped by $(\mathbb{Z}_2^{[0]})^{\rm 3d}_\mathrm{center}$ transformation, this transformation maps the local eigenphases as
\begin{align}
    \theta(\vec{x}) \mapsto \theta(\vec{x}) \pm \pi.
\end{align}
In fact, the distributions obtained numerically have a form $\rho(\theta) \approx \rho(\theta\pm\pi)$, implying that they respect the $(\mathbb{Z}_2^{[0]})_\mathrm{center}^{\rm 3d}$ symmetry, even though the deconfinement takes place at $\beta_\mathrm{H} \gtrsim 4.5$.\footnote{The behavior $\rho(\theta)\approx \rho(-\theta)$ reflects that the Polyakov loop matrix $P(\vec{x})$ is an $\SU(2)$ matrix and its two eigenphases are $\pm \theta(\vec{x})$.}
This situation is compatible with the fact that, in the finite system, no strict phase transition can be observed through the expectation value of one-point functions.

\begin{figure}[t]
    \begin{minipage}[b]{0.5\hsize}
        \centering
        \includegraphics[width=\textwidth]{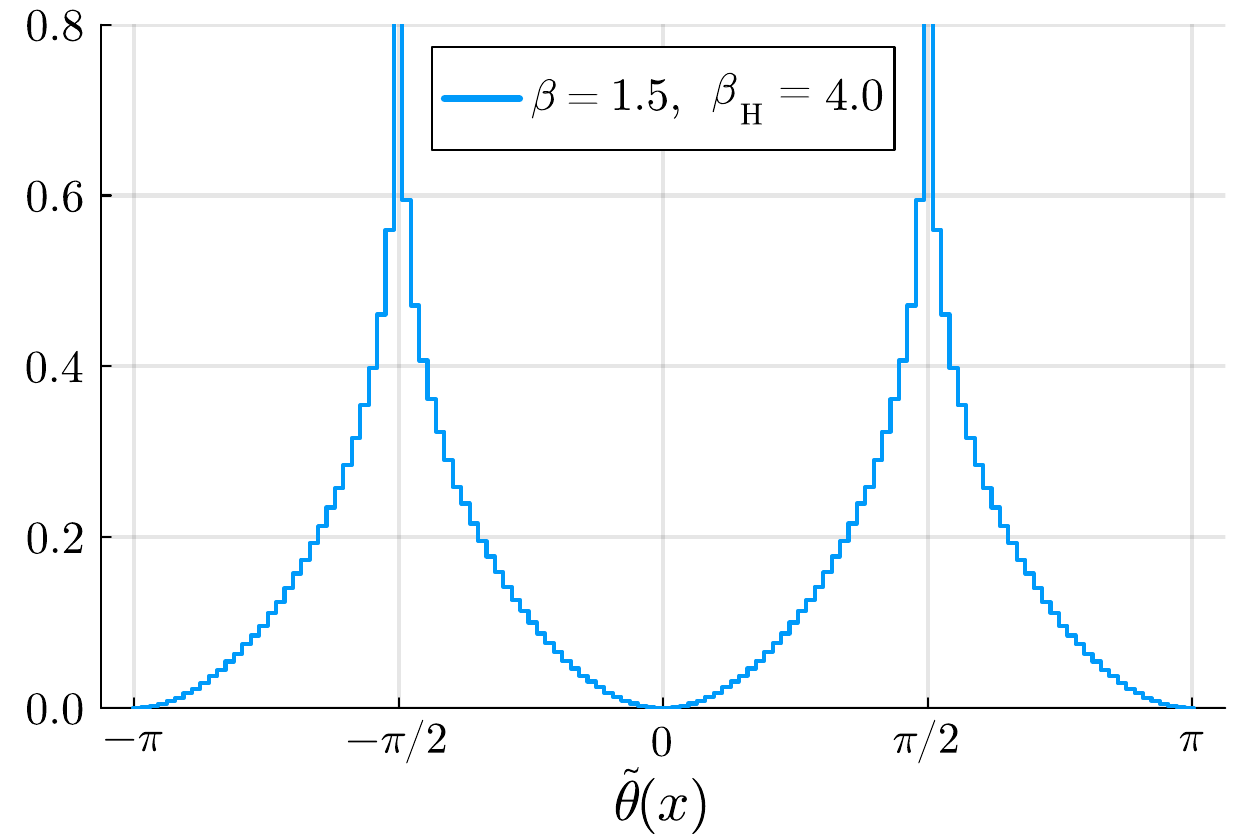}
        \subcaption{$\beta_\mathrm{H}=4.0$}
    \end{minipage}
    \begin{minipage}[b]{0.5\hsize}
        \centering
        \includegraphics[width=\textwidth]{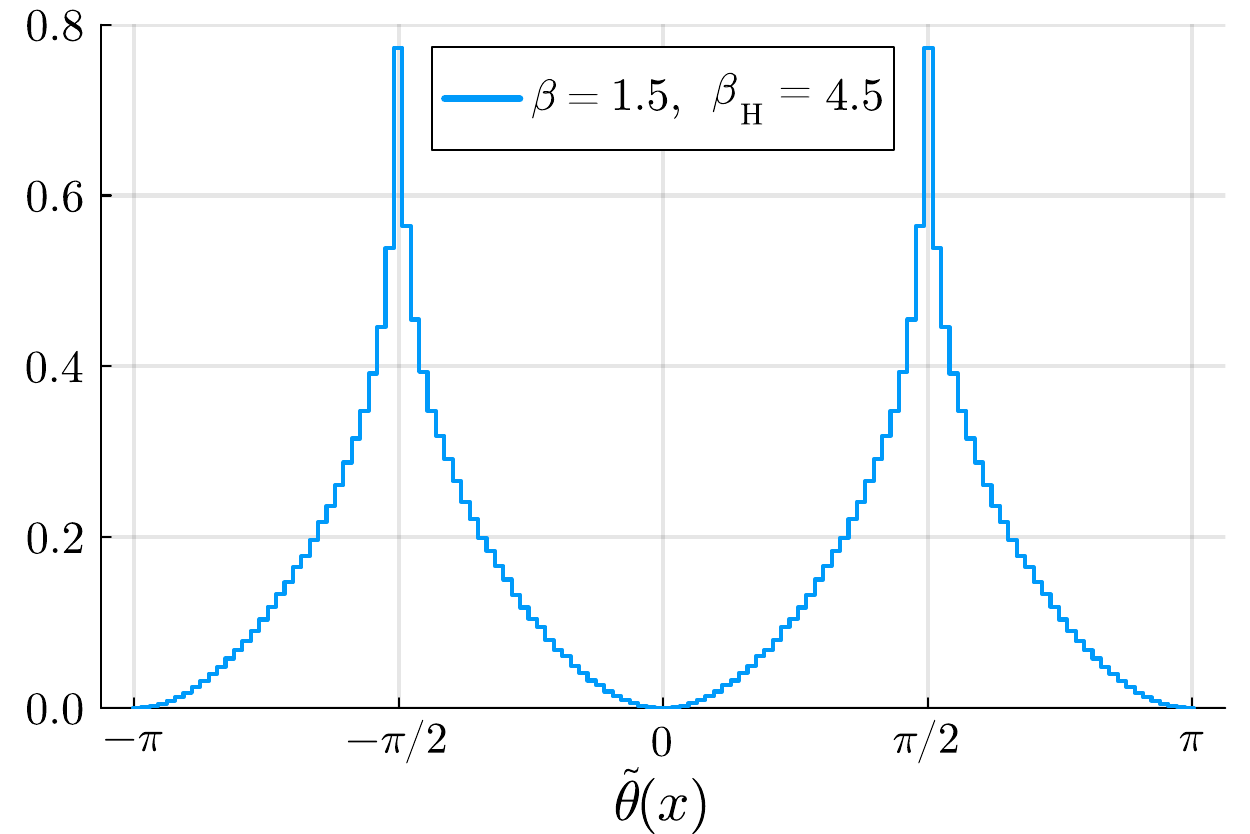}
        \subcaption{$\beta_\mathrm{H}=4.5$}
    \end{minipage}
    \begin{center}
    \begin{minipage}[b]{0.5\hsize}
        \centering
        \includegraphics[width=\textwidth]{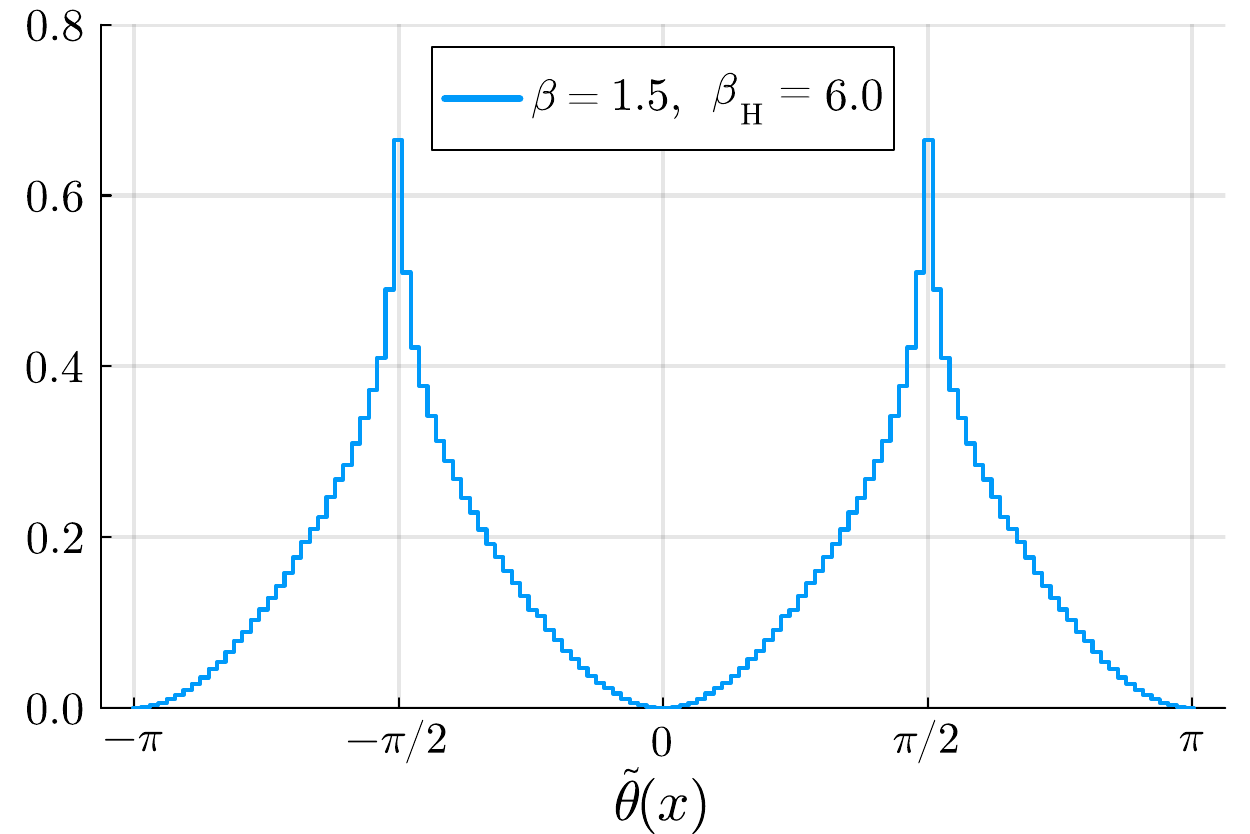}
        \subcaption{$\beta_\mathrm{H}=6.0$}
    \end{minipage}
    \end{center}
    \caption{
    Plots of the eigenphase distribution of $\im P^2\phi$ at $\beta = 1.5$ and several $\beta_\mathrm{H}$.
    }
    \label{fig:eigenphase_distrib_higgs_beta1.5}
\end{figure}

We can also introduce an eigenphase distribution of the $\SU(2)$ matrix $(\im P^2\phi)(\vec{x})$ in a similar manner.
By choosing a suitable $\tilde{V}\in \SU(2)$, we can obtain the eigenphase $\tilde{\theta}(\vec{x})$ of the matrix through  diagonalization as 
\begin{equation}
    (\im P^2\phi)(\vec{x})
    =
    \tilde{V}\diag(\ee^{\im\tilde{\theta}(\vec{x})},\ee^{-\im\tilde{\theta}(\vec{x})})\tilde{V}^{-1},
\end{equation}
and its distribution function as  
\begin{equation}
    \tilde{\rho}(\tilde{\theta})
    =
    \ev{\frac{1}{N_{\rm s}^3}\sum_{\vec{x}}\delta\big(\tilde{\theta}-\tilde{\theta}(\vec{x})\big)}.
\end{equation}
The $(\mathbb{Z}_2^{[0]})^{\rm 3d}_\mathrm{Higgs}$ symmetry is associated with a transformation that flips the sign of $\tr(\im P^2\phi)(\vec{x})$, which gives a mapping of eigenphases as 
\begin{equation}
    \tilde{\theta}(\vec{x}) \mapsto \tilde{\theta}(\vec{x}) \pm \pi.
\end{equation}

Figure~\ref{fig:eigenphase_distrib_higgs_beta1.5} plots the distribution function $\tilde{\rho}(\tilde{\theta})$ with $(N_\mathrm{s},N_{\rm t}) = (16,8)$ at fixed $\beta$ and various $\beta_\mathrm{H}$ values.
Note again that, in our numerical simulations, the adjoint Higgs field is fixed in the unitary gauge, $\phi \sim \sigma^y$.
These plots have a tendency $\tilde{\rho}(\tilde{\theta}) \approx \tilde{\rho}(\tilde{\theta}\pm\pi)$, which implies that the numerical simulation is performed while respecting the $(\mathbb{Z}_2^{[0]})^{\rm 3d}_\mathrm{Higgs}$ symmetry.

\subsection{Prospects on lattice simulations}
We conclude that the preliminary results of lattice simulations shown in this appendix are insufficient to settle the deconfinement-Higgs continuity of this lattice model.
There are mainly two reasons: the lack of simulation points and a long autocorrelation of observables.
Therefore, it is worthwhile to perform further lattice studies of this model to resolve them.

Regarding the first issue, a useful way to determine the position of the phase transition line and the possible critical point is to measure the susceptibilities. 
This method usually applies to the phase boundary between the confined and deconfined phases, utilizing the Polyakov loop susceptibility.
Simulations with different lattice sizes are therefore required to perform the finite-size scaling of susceptibilities.
Moving to a larger lattice is also beneficial from the aspect of studying correlation functions with a broader range of $r$.

Moreover, one must take care of the overwhelmingly long autocorrelation in the deconfined phase to improve the signal.
To achieve it, it would be better not to impose the unitary gauge in a large $(\beta,\beta_\mathrm{H})$ region. By making the adjoint field dynamical, an enhancement of tunneling is expected during the Monte Carlo sampling.
Practically, the presence of the dynamical adjoint fields enables us to use the so-called over-relaxation method for the link variables.

\newpage

\bibliographystyle{utphys}
\bibliography{./QFT,./refs,./kawahira}

\end{document}